\documentclass{aa}  

\usepackage{marginnote}
\usepackage{graphicx}
\usepackage{hyperref}
\usepackage{booktabs} 
\usepackage{siunitx}  
\usepackage{natbib}
\bibpunct{(}{)}{;}{a}{}{,}
\usepackage{multirow} 
\usepackage{caption}  
\newcommand{\Rs}{$ R_{\odot}$}

\newcommand{\de}{$^{\circ}$}

\newcommand{\km}{~km}
\newcommand{\kms}{~km$s^{-1}$}

\newcommand{\paperi}{Paper I}
\newcommand{\paperii}{Paper II}

\usepackage{txfonts}
\usepackage{booktabs}
\usepackage[table]{xcolor}

\begin{document}

   \title{A multi-viewpoint comparison of the velocity field of coronal propagating disturbances}

   \author{ Nina Stankovic, \inst{1} \thanks{E-mail: nis58@aber.ac.uk}
      Huw Morgan,\inst{1} \thanks{E-mail: hmorgan@aber.ac.uk}
  Marilena Mierla,$^{2,3}$
  Nancy Narang,$^{2}$ 
  Luciano Rodriguez$^{2}$\and
David Berghmans$^{2}$}
   \institute{Adran Ffiseg, Prifysgol Aberystwyth,
   	Ceredigion, Cymru, SY23 3BZ, UK
    \and 
    Solar-Terrestrial Centre of Excellence – SIDC, Royal Observatory of Belgium, Ringlaan -3- Av. Circulaire, 1180 Brussels, Belgium
     \and
  Institute of Geodynamics of the Romanian Academy, Bucharest, Romania}

   \date{Received October 06, 2025; accepted December 08, 2025}

  \abstract
   {Small-scale propagating disturbances (PD) are ubiquitous in the solar corona. Time-Normalised Optical Flow (TNOF) is a method developed for mapping PD velocity fields in time series of Extreme-Ultraviolet (EUV) images. We show PD velocity fields of a quiet Sun (QS) region containing a small coronal hole (CH) and filament channel (FC) jointly observed by Extreme Ultraviolet Imager (EUI) aboard the Solar Orbiter and Atmospheric Imaging Assembly (AIA) onboard the Solar Dynamics Observatory (SDO). The QS observations acquired on 28 October 2023 in 174\,\AA\ channel of  High Resolution EUV Imager (HRIEUV) of EUI and 171\,\AA\ channel of AIA are used. During the time of the observations, the separation angle between Solar Orbiter and SDO was approximately 26\de. A novel image alignment analysis shows that the dominant formation heights are 11.4\,Mm for HRIEUV and 4\,Mm for AIA. Despite this height difference, the PD velocity fields obtained from the observations from the two instruments are in good agreement across the region. In the QS the median PD speed is around 6.7 and 7.4\kms\ for HRIEUV and AIA respectively, with maximum speeds of around 40\kms. The small equatorial CH is a region dominated by a low temperature of $\approx$0.8MK and is host to high PD speeds, with a median speed of 17\kms. The velocity field bridges coherently across the CH from neighbouring QS regions from east to west, thus the CH must be overlaid by a system of long, low-lying closed magnetic loops. This unexpected configuration is supported by a potential field (PF) magnetic model and may be due to the longevity of the CH, allowing time for interchange reconnection with neighbouring closed-field regions. The FC is observed to be multithermal, with a narrow central strip of high emission at both low (0.8MK) and high (2.5MK) temperatures and low emission at warm (1.2MK) temperature. Despite this distinct temperature profile, the FC has PD speeds similar to those of the QS. The TNOF velocity field shows that PDs tend to flow into the FC from neighbouring regions before aligning along the FC in a coherent direction, thus PDs within filaments are driven by external sources. The vector field is consistent with a highly non-potential barbs-and-spine tubular magnetic field; a configuration that the PF model fails to replicate. We conclude that longer magnetic loops are required for higher PD speeds as observed for CH here, and that the smaller loop systems of the QS and FC generally lead to lower speeds. These multi-instrument results give high confidence in the TNOF method as a diagnostic tool for the kinematics of PDs, and highlights its potential for probing the coronal magnetic field orientation - particularly in highly non-potential regions where extrapolation models may fail. }


   \maketitle
%

\section{Introduction}

 The magnetic topology of the corona above the quiet Sun (QS) is only partly understood due to limited direct observations of coronal magnetic fields and a lack of constraints for magnetic field models in the atmosphere \cite[e.g.][]{khomenko_2003_quietsun, beck_2009_the, dorozcosurez_2012_analysis, anusha_2017_statistical}. Recent works have shown that the velocity fields of faint propagating disturbances (PD) viewed on-disk in extreme-ultraviolet (EUV) images can be used as an indirect tracer of the underlying coronal magnetic topology \citep{stankovic_2025_connections}. PDs play a significant role in plasma dynamics, facilitating energy and momentum transport while contributing to the heating and structuring of the solar atmosphere \citep{bellotrubio_2019_quiet, berghmans_2001_active}. In EUV time-series imagery, the PDs can be interpreted as signatures  of slow magnetoacoustic (MA) waves \citep{gupta_2012_spectroscopic, kolotkov_2021_the, banerjee_2021_magnetohydrodynamic, Longitudinal....156..123A, meadowcroft_2023_observation, barczynski_2023_slow, baweja_2025_plumes}, or as the high-speed quasiperiodic upflows \citep{tian_2011_the, schwanitz_2023_smallscale, 2023A&ASchwanitz}.

 Building on the work of \citet[][henceforth \paperi]{morgan_2022_tracing}, \citet{stankovic_2025_connections} showed that PDs are ubiquitous. By using Time-Normalised Optical Flow (TNOF) method, they observed that the coronal velocity fields were largely composed of a network of coherent cell-like structures, with their boundaries (PD sources) aligned with the photospheric supergranulation network. They found that PD sources were consistent across various temperature regimes, while sink regions showed weaker correlation and were often found in internetwork areas. The comparisons of velocity fields across the observations in various temperature channels showed source and sink regions in near-perfect alignment in many regions, and consistent spatial offsets in other regions. These were readily interpreted in terms of vertical and inclined magnetic field geometries by \citet{stankovic_2025_connections}, supporting the use of PD velocity maps to understand the topology of magnetic fields.
 
 This paper aims to study the velocity fields of a QS region containing a filament and equatorial coronal hole, and compares results from the Atmospheric Imaging Assembly (AIA, \citealt{lemen_2012_the}) onboard the Solar Dynamics Observatory (SDO, \citealt{pesnell_sdo}) and the Extreme Ultraviolet Imager (EUI, \citealt{rochus_2020_the}) onboard Solar Orbiter (SO, \citealt{mller_2020_the}). The High Resolution EUV Imager (HRIEUV) of EUI observing in the 174\AA\ channel offers a unique opportunity to study small-scale solar activity including, but not limited to, small-scale brightenings \citep{berghmans_2021_extremeuv, narang_2025_euvbrightenings, 2025Lim}, and pico-flare jets \citep{chitta_2021_capturing, 2023SciChitta} in the solar corona.
 
 The region of interest presented in this study is QS containing a quiescent filament channel (FC) and a small transequatorial coronal hole (CH) \citep[e.g.][]{drielgesztelyi_2006_magnetic}. QS filaments are generally larger than those in active regions, with the elongated central dark core, or spine, ranging from 60 to 600Mm. They lie above polarity inversion lines in the photosphere, and consist of numerous fine threads aligned with the local magnetic field, with lateral offshoots called barbs \citep{baso_2019_spectropolarimetric, diercke_2018_counterstreaming, ckuckein_2016_giant}. When viewed above the limb, they are called prominences, and are often surrounded by a large hot cavity \citep{Habbal_2010}. The overall filament-cavity structure is thought to be a large magnetic flux tube with the cooler plasma condensing at the base of the tube \citep{Hutton_2015}. Filaments exhibit oscillations and wave-like motions, often linked to magnetoacoustic waves, which can propagate along their structures, trigger mass flows, and cause periodic disturbances observed in Doppler velocity measurements \citep{somaiyeh_2024_magnetoacoustic}. 
 
 Compared to QS regions, equatorial CHs are characterised by a generally lower overall emission in the EUV, indicative of lower density, and their emission measures peak at lower temperatures \citep{morgan_2017_global, morgan_pickering_2019}. Coronal holes, particularly large polar CHs, are usually characterised by open, unipolar magnetic fields \citep{wang_2008_coronal} that extend into the heliosphere, distinguishing them from the mixed-polarity fields of the QS. The topology of coronal holes can rapidly evolve due to interactions with nearby active region or QS magnetic fields, a process influencing their unipolar magnetic configuration \citep{lezzi_2023_dark, cranmer_2009_coronal, vandriel2012}. On smaller scales, coronal holes can show a more complex magnetic field configuration characterised by mixed polarities \citep{twiegelmann_2004_similarities, schwanitz_2023_smallscale,huang_2012_coronal}. Equatorial CHs have a weak photospheric field of mixed polarity but with one polarity slightly dominating when integrated over the whole region (e.g. \paperi). \citet{eoshea_2006_a} found evidence of propagating slow magnetoacoustic waves in polar and as well as in equatorial coronal holes. Their results show MA waves observed in coronal holes \citep{banerjee_2009_propagating} contribute to non-thermal line broadening in EUV spectral lines. \cite{harra_2025_the} provides overview of recent advancements in understanding the dynamics of coronal holes and quiet Sun.
 
 This paper presents an in-depth comparative analysis of PD velocity fields within different magnetic sub-regions of the field-of-view observed by HRIEUV, namely QS, equatorial CH and quiescent filament channel (FC). The paper is structured as follows: section 2 presents the HRIEUV and AIA observations and methodology. Section 3 presents results of vector field maps, statistical speed distributions in various regions, interpretation in terms of magnetic topology, and a comparison between the two instruments. In section 4, we further discuss the  results and section 5 presents conclusions with ideas for future work.\\
 
 \section{Data set and methodology} \label{sec:2}
\subsection{Observation Details}

\begin{figure*}
	\centering
	\includegraphics[width=0.45\textwidth]{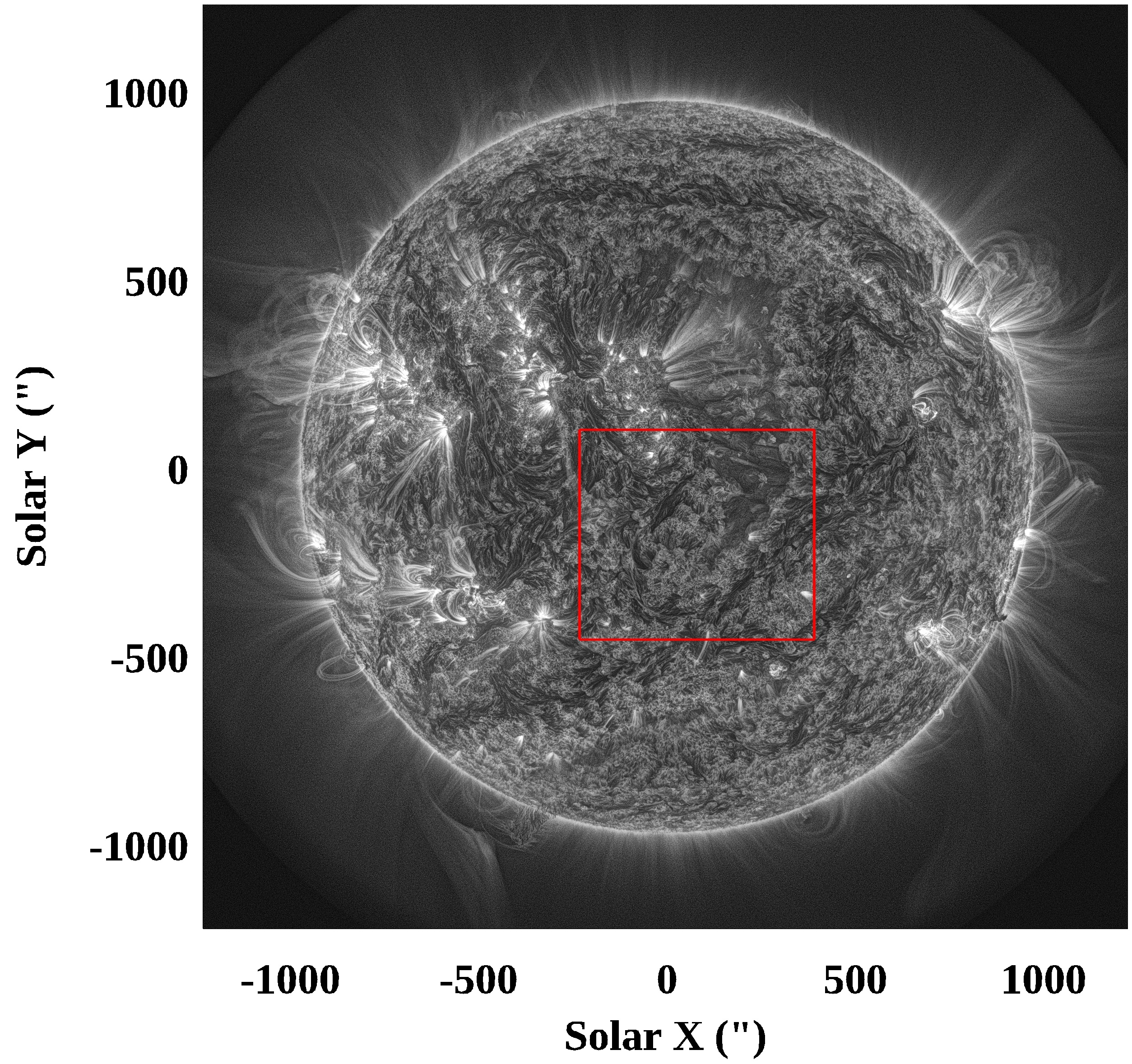}
	\includegraphics[width=0.45\textwidth]{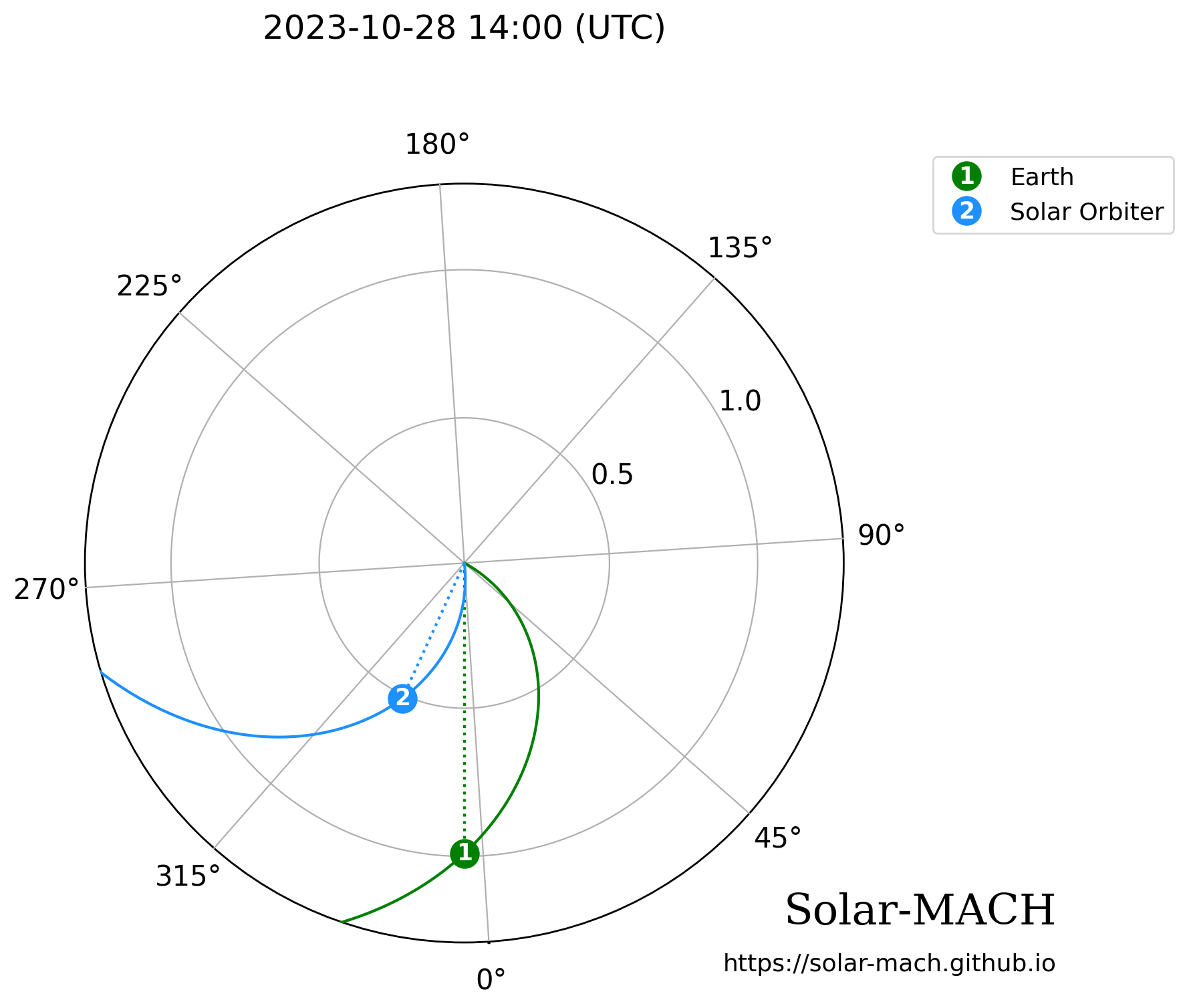}
	\caption{Left: AIA 171\r{A} full-disk image on 2023 October 28 14:04UT with the ROI boxed in red. This image has been processed using Multiscale Gaussian Normalisation (MGN). \protect\citep{morgan_2014_multiscale} Right: Illustration from the Solar-MACH tool \protect\cite{gieseler_2023_solarmach} 
		where the location of the Earth/SDO marked by (1) and Solar Orbiter marked by (2) relative to Sun is shown.}
	\label{6.0}
\end{figure*}

The two datasets used in this work are observations taken by HRIEUV at 174 \r{A} and AIA at 171 \r{A} on 2023 October 28. The region of interest (ROI) is mostly QS, centred just southwest of disk centre, as shown by the red bounding box in the left panel of Fig.\ref{6.0}. This unique dataset is currently one of the longest continuous observations of QS obtained by HRIEUV, with high temporal and spatial resolution. The HRIEUV observations span a two-hour interval from 14:00UT until 16:00UT, with a cadence of five seconds. At this time, SO was located at a distance of 0.52AU from the Sun. From this distance, the pixel scale of 0.492$\arcsec$ of HRIEUV images \citep{gissot_2023_initial} correspond to one-pixel resolution of approximately 200\km on the solar surface. Thus the $2048 \times 2048$ pixel images correspond to a FOV of  $410\times410$~Mm on the Sun's surface. AIA and HRIEUV simultaneously captured this region from two different perspectives, at a longitudinal separation angle of around 26\de\ as depicted in the right panel of Fig.\ref{6.0}. The AIA observations have a cadence of 12 seconds with a pixel size of $0.6\arcsec$ observing at a distance of 1AU from the Sun.

The Level-2 calibrated data of HRIEUV is used (with pre-processing applied to reduce spacecraft pointing errors and jitter in the images) from EUI data-release 6 \citep{euidatarelease6}. The specific HRIEUV observation sequence used in this work was obtained with Solar Orbiter Observing Plan (SOOP) called 'R\_BOTH\_HRES\_MCAD\_Bright-Points'\citep{auchre_2020_coordination,zouganelis_2020_the}. For AIA observations, We used calibrated level 1 scientific data from the AIA instrument that is available from the Joint Science Operations Centre  (\citeauthor{a2025_joint}), and applied the standard AIA data-processing steps using SSWIDL routine of \textit{aia\_prep.pro}.    

\begin{figure*}
	\centering
	\includegraphics[width=0.96\textwidth]{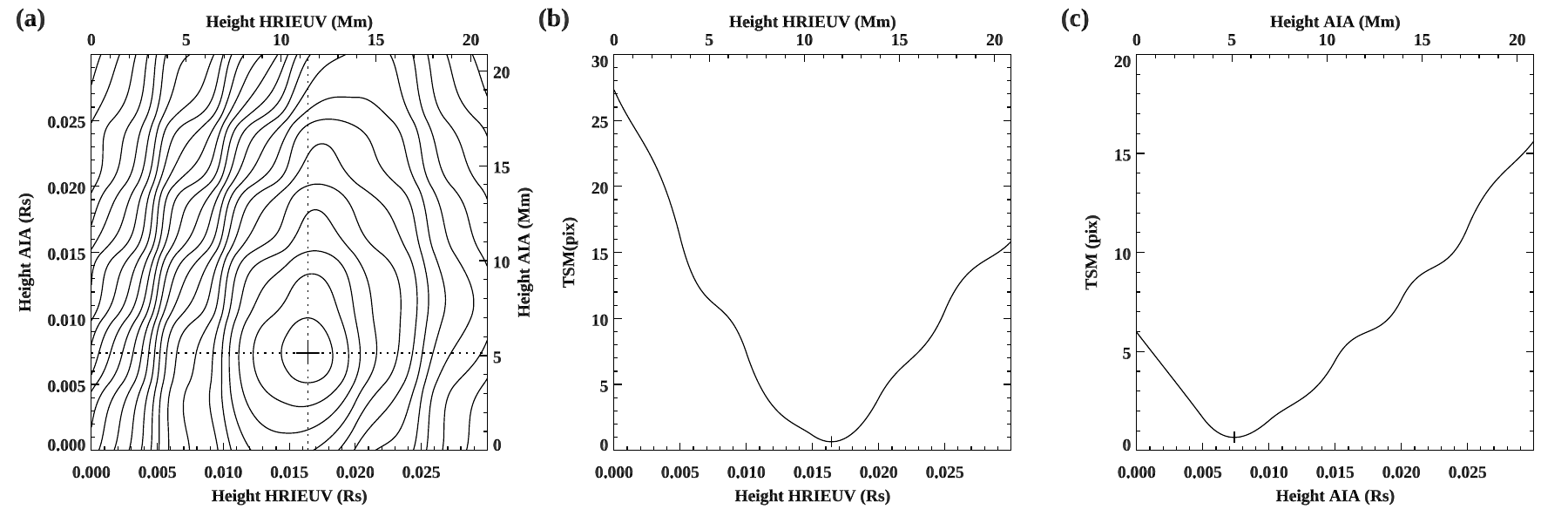}
	\caption{(a) TSM or the magnitude of the translational pixel shift between the HRIEUV and AIA images after remapping to Carrington longitude-latitude coordinates, as a function of height for HRIEUV ($x$-axis) and AIA ($y$-axis). These heights describe the surface used to map to spherical coordinates, and are shown in both \Rs\ and Mm. The cross shows the minimum of the surface. (b) and (c) show TSM as a function of height for cuts passing through the minimum point, shown by the dotted horizontal and vertical lines in (a).}
	\label{euiaiageom}
\end{figure*}

In order to compare results from both viewpoints we transform all images into Carrington coordinates. This presents a challenge since the transformation requires a mapping to a spherical surface. When we use the photosphere as the surface, a consistent shift is seen between features in HRIEUV and AIA, since the observed light is dominated by emission from the low corona above the photosphere. We therefore experiment with surfaces at different heliocentric distances, ranging from the photosphere into the corona. The images taken closest to the middle time of our period of interest (14:05UT) are used for both instruments. We vary the distance independently for both instruments since we assume their emission may be dominant at different heights. For each pair of heights (HRIEUV height, and the independent AIA height) we remap the images into a regular Carrington longitude and latitude grid projected to that spherical surface, and calculate the translational $\Delta x$ and $\Delta y$ sub-pixel shifts required to maximise the spatial alignment between the images using Fourier Correlation Tracking \citep[FCT,][]{fisher_2008_flct}. This analysis is limited to the region of interest defined by the HRIEUV FOV, given that the AIA observes the full solar disc. For each pair of images we calculate a Translational Shift Magnitude as TSM$=\sqrt{\Delta x + \Delta y}$. Figure \ref{euiaiageom}(a) shows how TSM varies as a function of height. The calculations are made at increments of $\approx$1.5Mm, and the results of Figure \ref{euiaiageom} show the cubic spline interpolation of the discrete results. This surface reaches a minimum TSM of $\approx$1 pixel at a height of 11.4Mm for HRIEUV and 4 Mm for AIA. Cuts of the TSM as a function of height are shown in Figures \ref{euiaiageom}(b) and (c) for HRIEUV and AIA respectively. These optimal distances define our spherical surface for remapping into Carrington coordinates for all our subsequent results, and enable the best comparison of features across the ROI. 

The above described method for minimising the translational shift between images uniquely gives the approximate dominant height of emission of the 171 and 174\AA\ channels of AIA and HRIEUV respectively. These optimal heights are only approximated average heights, since the emission comes from a range of heights. The height of maximum emission can also vary across the region, particularly between and even within different coronal structures (i.e. QS, CH, filament channel). The optimal heights are therefore an overall average estimate that is suitable for the whole ROI that may lead to local differences.This can likely be the reason why the TSM shown in figure \ref{euiaiageom} has a minimum of $\approx$one pixel rather than zero. Applying this method for smaller regions in order to map the dominant height of emission in different structures leads to less stable curves than that for the whole region, and requires further development.

In simplified terms, the separation between the two spacecraft is 26 degrees, which means that a feature extending 4 Mm above the surface introduces a parallax displacement of about 1753 km. With AIA’s rebinned pixel size of 864.6 km, this corresponds to about 2 pixels. Since AIA’s effective spatial resolution is around two pixels (about 1730 km), the parallax uncertainty that remains within the AIA spatial resolution. Because the source and sink regions are expected to lie in the low corona, the optimised height assumed in the alignment is biased towards the altitude of strongest emission. The large-scale structure remains unchanged over the time period examined, so any artificial uncertainty introduced by height differences would be static rather than dynamic. 

\begin{figure*}
	\centering
	\includegraphics[width=1\textwidth]{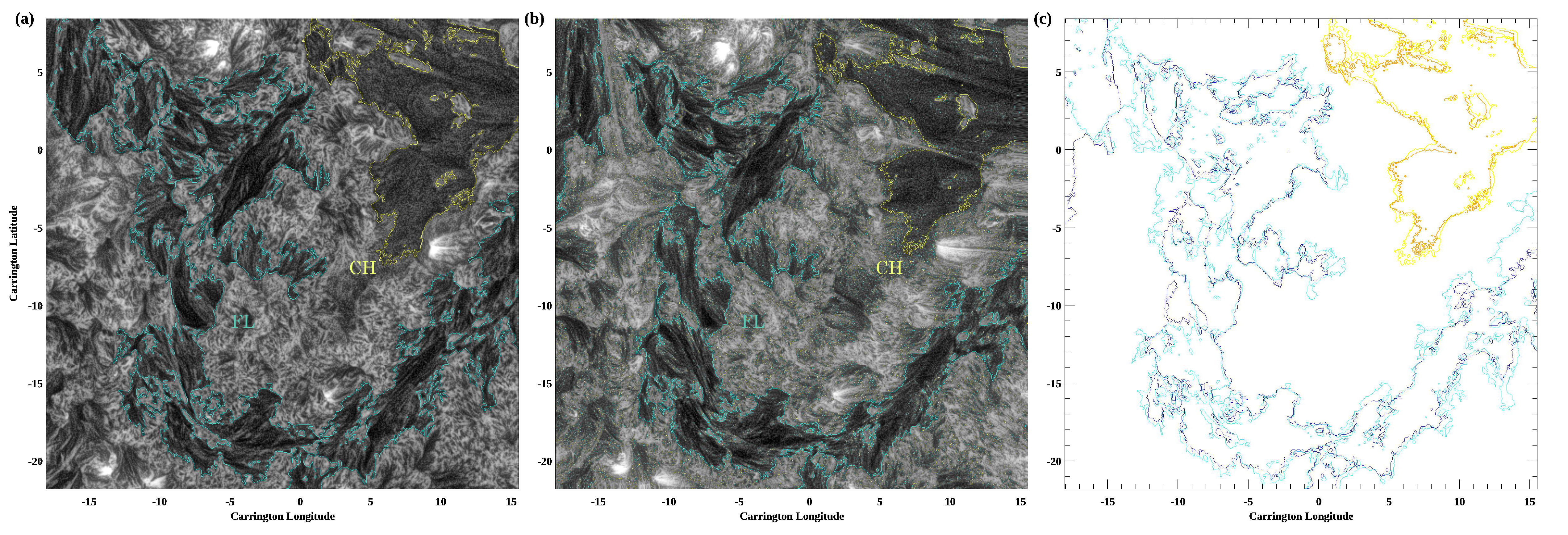}
	\caption{The ROI as observed by (a) HRIEUV  and (b) AIA. For display purposes, these images have been processed using MGN. The images are mapped into Carrington coordinates to better compare features observed from the two viewpoints. The yellow contour bounds a transequatorial CH, and the filament is labelled with a cyan contour. (c) A comparison of the segmented filament and CH regions from both instruments with the filament in dark blue contours (HRIEUV) and cyan (AIA), and the CH in orange (HRIEUV) and yellow (AIA).}
	\label{6.1}
\end{figure*}

Following remapping to Carrington coordinates, Figure \ref{6.1} shows a closer view of the ROI for (a) HRIEUV and (b) AIA, with the CH bounded by a yellow contour, and the filament(s) in cyan. These regions are identified by labelling all pixels below a certain intensity threshold, and keeping only continuous regions of area larger than $2\times 10^4$ pixels. The thresholds differ between the instruments, and there is a small difference in thresholds between the CH and filaments allowing their distinction. These thresholds are found by manual inspection and trial-and-error. Figure \ref{6.1}(c) superimposes the filament and CH contours from both instruments, with the CH in orange and yellow, and the filament(s) in dark blue and cyan for HRIEUV and AIA respectively. 

\begin{figure*}
	\centering
	\includegraphics[width=0.9\textwidth]{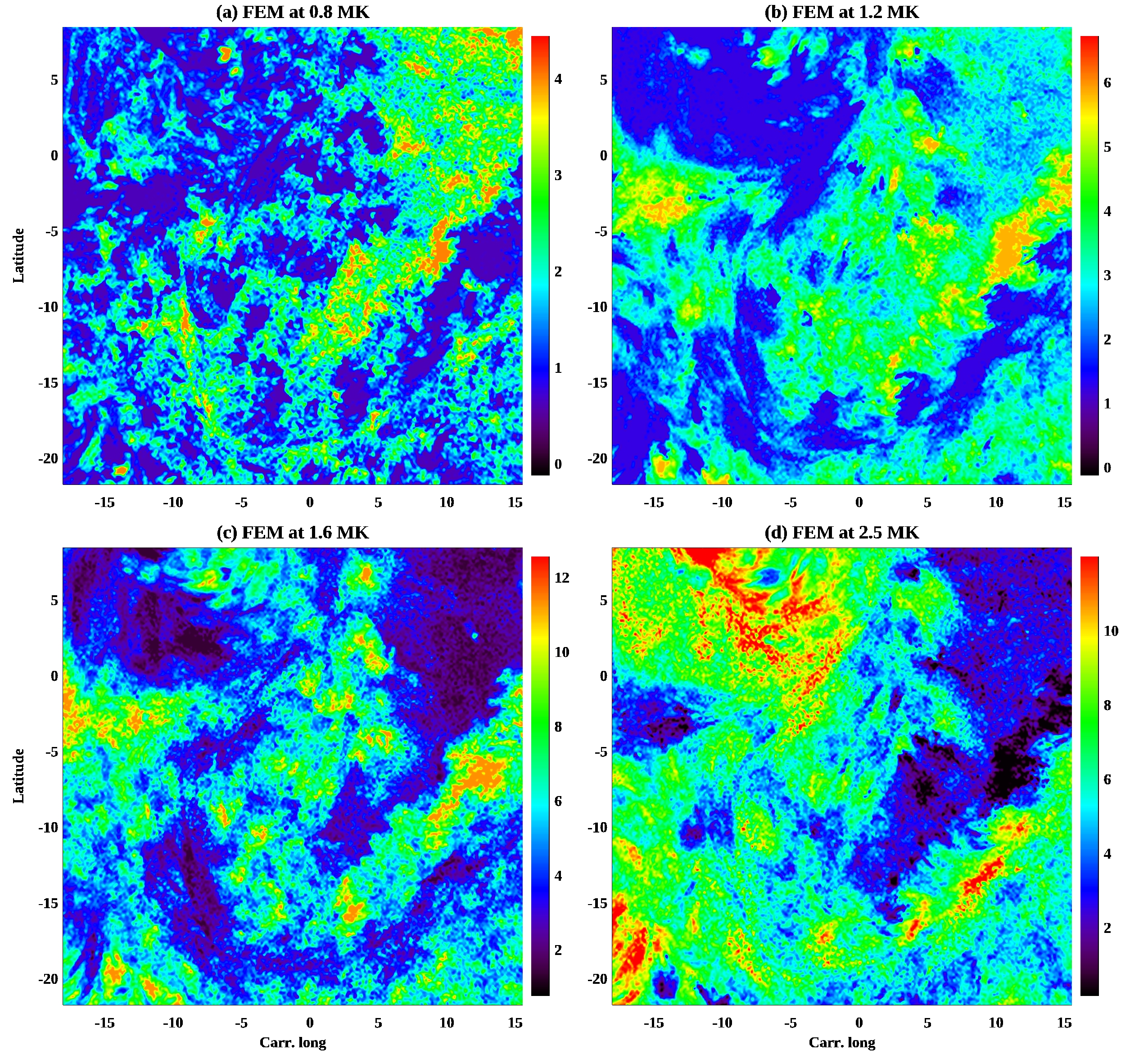}
	\caption{Results of a Differential Emission Measure (DEM) analysis of the ROI at four selected temperatures (a) 0.8, (b) 1.2, (c) 1.6, and (d) 2.5 MK. These plots show the Fractional Emission Measure (FEM), which is the emission at a given temperature divided by the total emission (integrated over all temperatures) at that pixel expressed as a percentage as shown in the colour bars.}
	\label{dem}
\end{figure*}

It is often difficult to distinguish small CH regions near the equator from other types of regions, such as dark halo regions \citep[e.g.][]{lezzi_2024_first}. The CH has been reported as such in the Heliophysics Event Knowledgebase (HEK) based on the Spatial Possibilistic Clustering Algorithm \citep[SPoCA,][]{cisverbeeck_2014_the}, and seems to be a southward equatorial extension to a larger CH to the north. This northern CH and it's equatorial extension is clear as a dark region in AIA 193\AA\ images. This CH exists for several rotations prior to our date and time of study, and there are no identified active regions near the ROI for a similar time period. Figure \ref{dem} provides results from a Differential Emission Measure (DEM) analysis to the ROI. We use all EUV channels of AIA for this analysis, and apply the Solar Iterative Temperature Emission Solver (SITES) algorithm for the inversion \citep{morgan_pickering_2019}. To better delineate regions of different temperature, we show here the Fractional Emission Measure (FEM), which, for a given pixel, is the emission at a temperature divided by the total emission integrated over all temperatures, expressed as a percentage. At a low temperature of 0.8MK, Figure \ref{dem}(a)shows the CH as a region of relatively high FEM. The southern filament channel has a narrow central line of relatively high FEM at this temperature. At 1.2MK (Figure \ref{dem}(b)), the CH FEM has dropped and the boundaries of the CH have a higher FEM. The filament has very low FEM at this `warm' temperature. At higher temperatures of 1.6MK and above (Figure \ref{dem}(c) and (d)) the CH has very low FEM. The filament has a wide strip of high FEM at the hot temperature of 2.5MK which we interpret as hot plasma surrounding the cooler filament itself which is supportive of eclipse observations of hot prominence cavities \citep{habbal_2010_total}. In summary, the identification of the north-west part of the ROI as a CH is supported by this temperature analysis, its persistence over several rotations, and it's association with a larger CH in the north as described above.

\subsection{Time-normalised optical flow (TNOF)}

The TNOF method works in two main steps, summarised here with full details given by \cite{MorganKorsos2022} (henceforth \paperii). First, small-scale dynamics including faint PDs are enhanced through a time-normalisation process. The second step involves an optical flow algorithm to characterise any quasi-repetitive faint motions. 

A time series of images of the ROI are loaded into a data cube, and the individual frames are globally co-aligned via sub-pixel translations found using Fourier correlation to a central reference image taken closest to the mid-time of the observation period \citep{fisher_2008_flct}. Following alignment, the images are spatially rebinned through local averaging, discussed further below. High-frequency noise is then reduced by applying a narrow Gaussian low-pass filter to each pixel's time series. The time series is filtered by subtracting a time-local weighted mean (calculated by convolution with a wide Gaussian kernel) and dividing by the local weighted standard deviation (calculated over the same wide kernel) plus a small constant, which produces a filtered signal with a local zero mean and a standard deviation of one - this is the time-normalisation step. The small constant avoids dividing by very small numbers in low-signal regions. Additional noise reduction is achieved through a spatial multiscale \textit{á trous} decomposition that discards the highest frequency component and sums the remaining scales (applied to each time frame independently). 

In the second step, a modified Lucas-Kanade optical flow algorithm is used to derive a velocity vector field, capturing both the direction and magnitude of motion across the region of interest. The algorithm calculates the spatial ($x$ and $y$) and temporal ($t$) derivatives of the intensity, and finds a least-squares solution for the velocities $v_x$ and $v_y$ at each pixel using the standard relationship between the derivatives on which Lucas-Kanade is based (see equation 4 of \paperii). Since we have 600 time steps at each pixel for AIA, and 1400 for HRIEUV, the least-squares solution can be found locally at each pixel independently (the Lucas-Kanade algorithm usually operates on pairs of images, and thus requires a local group of pixels for a solution). Shorter observation periods lead to less coherent velocity fields due to fewer data points, causing increased fluctuations at small scales and decreased correlation with the coronal structure. The velocities (in units of pixels per time step) are used to warp the image intensities, essentially reducing any motion shifts between time steps, and the process is repeated. Iteration continues until the change in the velocity field becomes small. There are further steps at each iteration including imposing an adaptive local spatial smoothing on the velocities based on the fitting residuals, all detailed in \paperii. Based solely on the fitting residuals, \paperii\ gives an uncertainty of 6.5\%\ on the derived velocities for a QS region in the 193\AA\ AIA channel. This should be considered a minimum estimate of uncertainty.

Once velocities are found, the units are converted from pixels per time step to \kms\ using the appropriate spherical coordinates of each pixel (based on the optimal heights given by the TSM approach described above). It's important to keep in mind that these are the plane-of-sky `optical velocities' that can suffer from projection effects. We minimise this by choosing a region near the disc centre, and by the appropriate use of spherical coordinates at the optimal height. Changing the width of smoothing kernels result in different velocity magnitudes although the vector directions remain consistent (see section 5 of \citet{morgan_2018_ubiquitous}). \citet{morgan_2018_ubiquitous} also shows a bias in the $v_x$ values consistent with the solar rotational velocity which gives some confidence that the optical velocities are reasonably well calibrated. In this work, the step of co-aligning the datacube images removes this bias.

 \begin{table}
 	\centering
 	\caption{Comparison of mean, median, and maximum speeds for HRIEUV and AIA for varying rebin factors.}
 	\label{tab:rebin_speeds}
 		\small 
 		\begin{tabular}{cccccc}
 			\toprule
 			\multirow{2}{*}{Inst.} & \multirow{2}{*}{R} & \multicolumn{3}{c}{Speed (km/s)} \\
 			\cmidrule(lr){3-5}
 			& & Mean & Median & Max \\
 			\midrule
 			\multirow{5}{*}{HRIEUV} & 2 & 6.90 & 5.62 & 79.06 \\
 			& 3 & 8.40 & 6.72 & 93.38 \\
 			\rowcolor[gray]{0.8} HRIEUV & 4 & 9.64 & 7.67 & 98.70 \\
 			& 5 & 10.72 & 8.56 & 141.18 \\
 			& 6 & 11.64 & 9.34 & 163.78 \\
 			\midrule
 			\rowcolor[gray]{0.8} &2 & 8.95 & 7.10 & 42.87 \\
 			& 3 & 10.80 & 8.39 & 61.41 \\
 			AIA& 4 & 12.13 & 9.37 & 66.46 \\
 			& 5 & 13.03 & 10.17 & 71.40 \\
 			& 6 & 13.78 & 10.78 & 62.70 \\
 			\bottomrule
 		\end{tabular}
 \end{table}
For computational efficiency and improved noise statistics, we spatially rebin images prior to applying the TNOF method. This is a rebinning using local averaging to a new image size that is an integer division of the number of original pixels. Here, for comparing the results from the two instruments with different spatial pixel scales, we can also use this rebinning step to better match the pixel spatial scales of both instruments. The closest matching pixel scales are given by integer rebinning factors of two for AIA and four for HRIEUV. For HRIEUV sequence used here, the observed solar radius is 1851.90\arcsec, and the pixel scale is 0.49\arcsec\ which is 184.8\km\ at the Sun. For AIA, the observed solar radius is 965.88\arcsec, and the pixel scale is 0.60\arcsec\ which is 432.3\km. Therefore, the rebinned images have a pixel scale of 739.32\km\ for HRIEUV, and 864.6\km\ for AIA. Table 1 shows how different rebinning factors affect TNOF speed values. The mean, median and maximum values of the speeds across the ROI are listed for HRIEUV and AIA for rebinning factors from 2 to 6. The highlighted rows show our final selection for rebinning values based on comparable pixel size. Using a higher rebinning value gives a smoother velocity field and higher speed values. When comparing the same bin factors between instruments, we see that AIA always has higher speeds because of the larger pixels.

Our results assume a similar measurement of the plasma by both instruments. \cite{shestov_2025_crosscal} compared the temperature response of HRIEUV and the AIA channels (see, e.g., their Figure 2 and 3) and showed that the radiometric sensitivity of HRIEUV is approximately 40\% more than that of AIA 171\AA. The HRIEUV bandpass is reported to be slightly broader than AIA 171\AA  bandpass \cite{shestov_2025_crosscal, adolliou_2023_temperature}, with the HRIEUV bandpass centred very close to the Fe IX/X emission and have peak temperature response at 1.0MK. The AIA 171\AA\ bandpass is dominated by Fe IX which peaks closer to 0.8MK. HRIEUV is therefore dominatingly more sensitive to a slightly hotter plasma than AIA 171\AA. This is supported by our TSM analysis, where the optimal height for HRIEUV is considerably higher, and thus hotter, than for AIA 171\AA. The large height difference, yet relatively small temperature difference, implies a shallow temperature increase gradient in the low corona for this region.

\section{Results}

\subsection{Velocity vector fields}

\begin{figure*}
	\centering
	\includegraphics[width=0.49\textwidth]{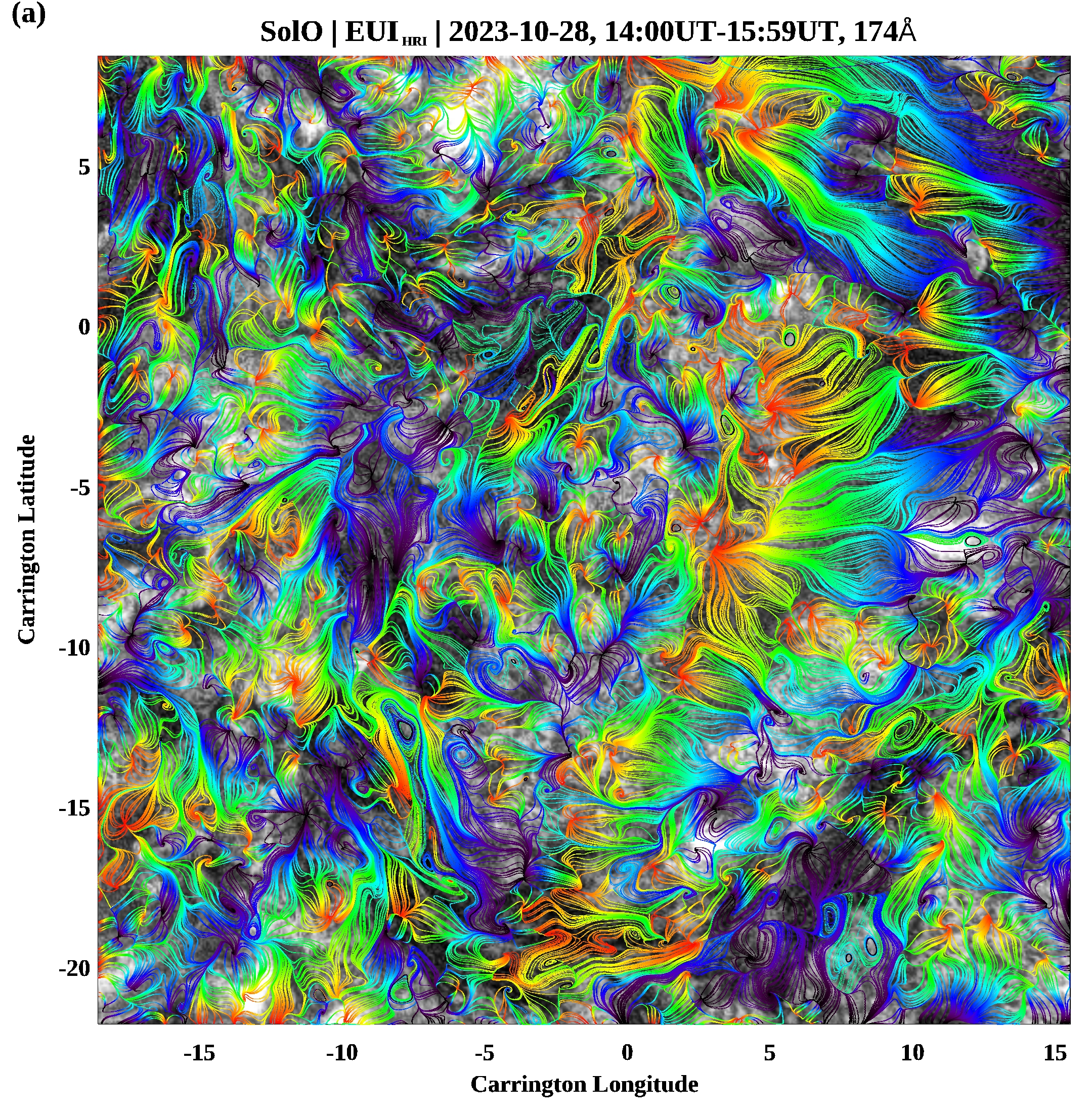}
	\includegraphics[width=0.49\textwidth]{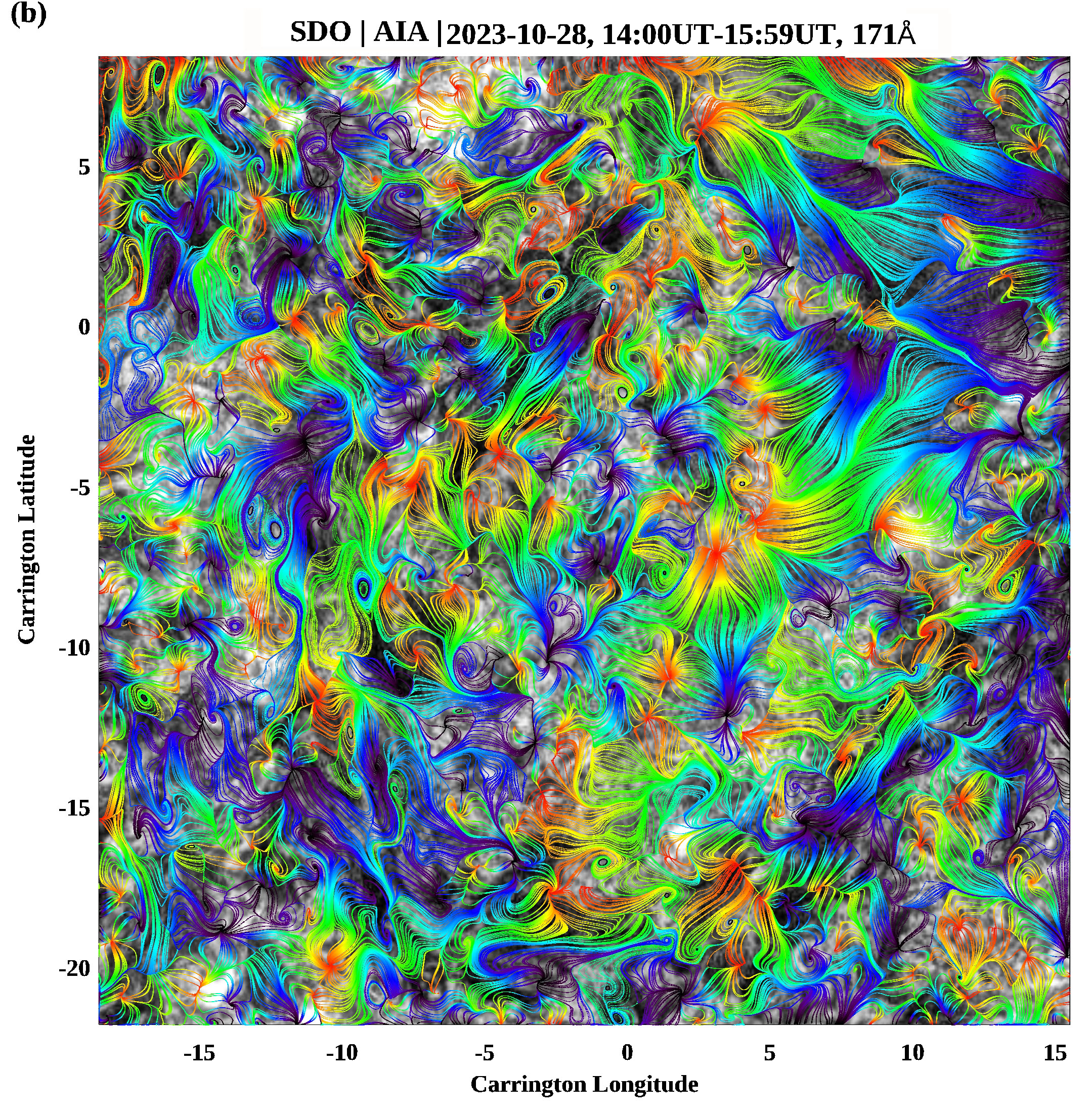}
	\caption{The TNOF velocity vector field for the (a) HRIEUV 174\r{A} and (b) AIA 171\r{A} channel. The fieldlines have a rainbow colour scheme to indicate propagation direction, with a field line starting with the colour red, and advancing through yellow, green, and ending in blue. The background image is the original intensity image with MGN processing.}
	\label{6.2}
\end{figure*}

The velocity fields, superimposed on the background intensity images, are shown in Figure \ref{6.2} for HRIEUV and AIA. For visualization, fieldlines are plotted with a rainbow colour scheme, from red, yellow, green to blue, to indicate propagation direction, with a field line starting with the colour red, and advancing through yellow, green, and ending in blue. The vector maps show similarities in both channels across most of the ROI, with good correspondence between larger-scale coherent structures. Differences can arise due to projection effects inherent to the different longitudinal viewpoints, differences due to assuming a constant height in mapping to Carrington coordinates, inherent differences in the data (pixel-size, cadence, noise).Some differences may also appear because of the display scheme of the vector field, i.e. the fieldlines are traced from a random distribution of points from the TNOF results of the two nearly aligned data-sequences. The different temperature response of the two instruments will also lead to differences. 

Within the QS (outside of the filaments and CH), the field is typically distributed as a network of distinct cells, many circular/elliptical with a typical diameter of 30- 50\arcsec. There are other more elongated features, with coherent sets of fieldlines extending across several degrees of longitude/latitude. Many of these structures originate from a concentration of PD sources (red sections of lines), but there are also concentrations of PD sinks (blue/black sections). Long spines of converging or diverging flows often connect multiple source or sink regions, with clear, narrow boundaries marked by very low or zero velocities. 

\begin{figure*}
	\centering
	\includegraphics[width=0.4\linewidth]{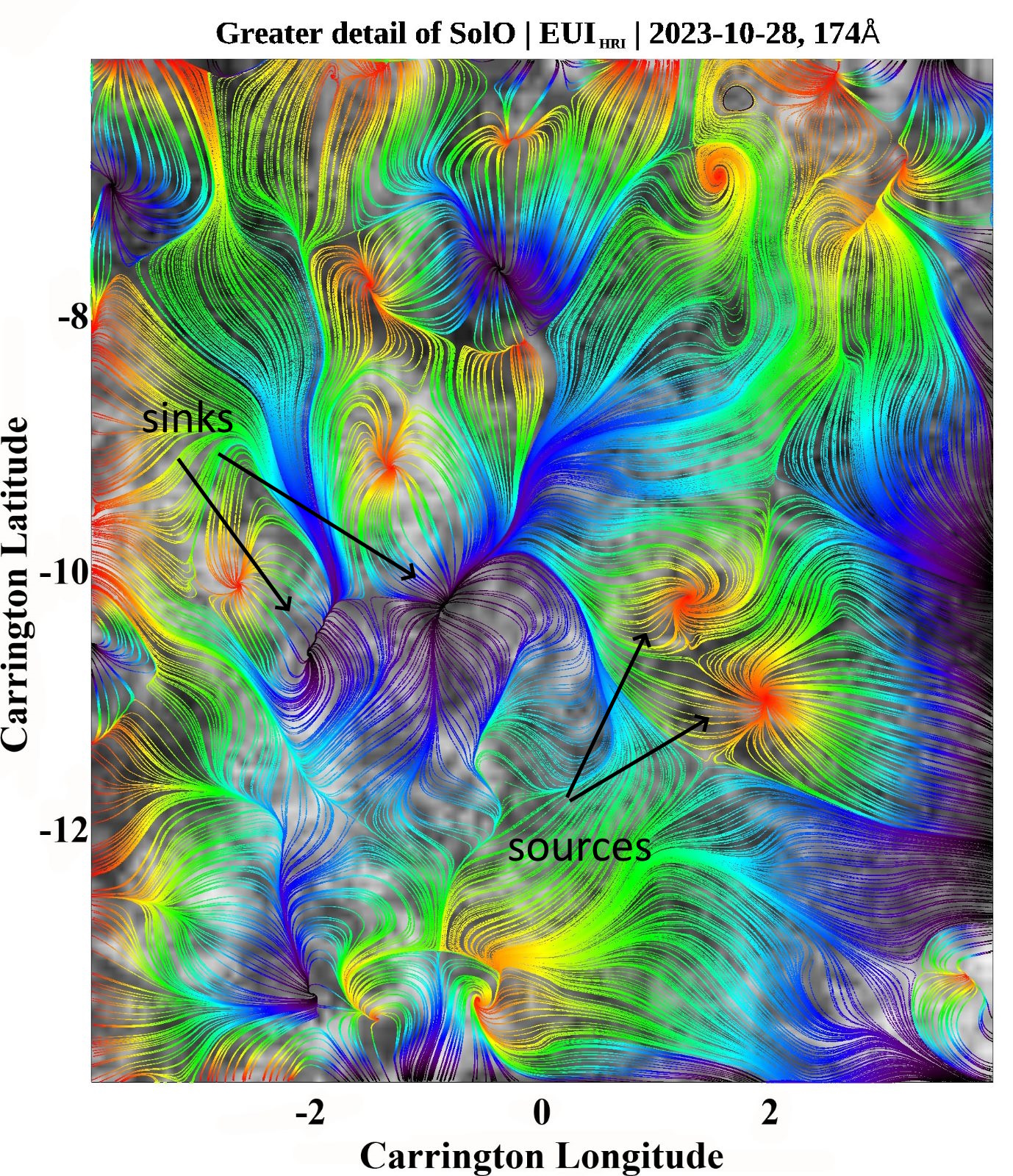}
	\includegraphics[width=0.4\linewidth]{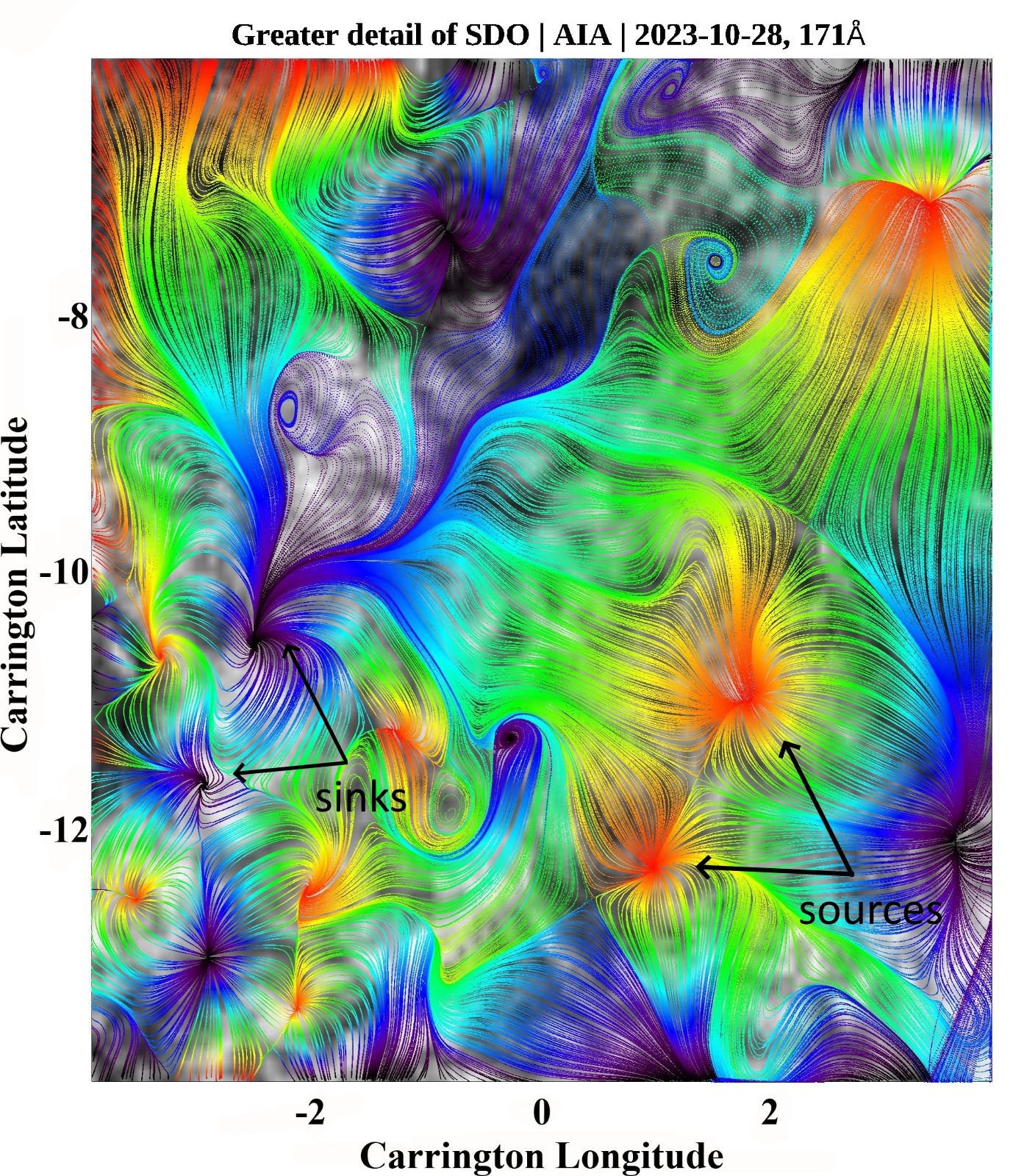}
	\includegraphics[width=0.45\linewidth]{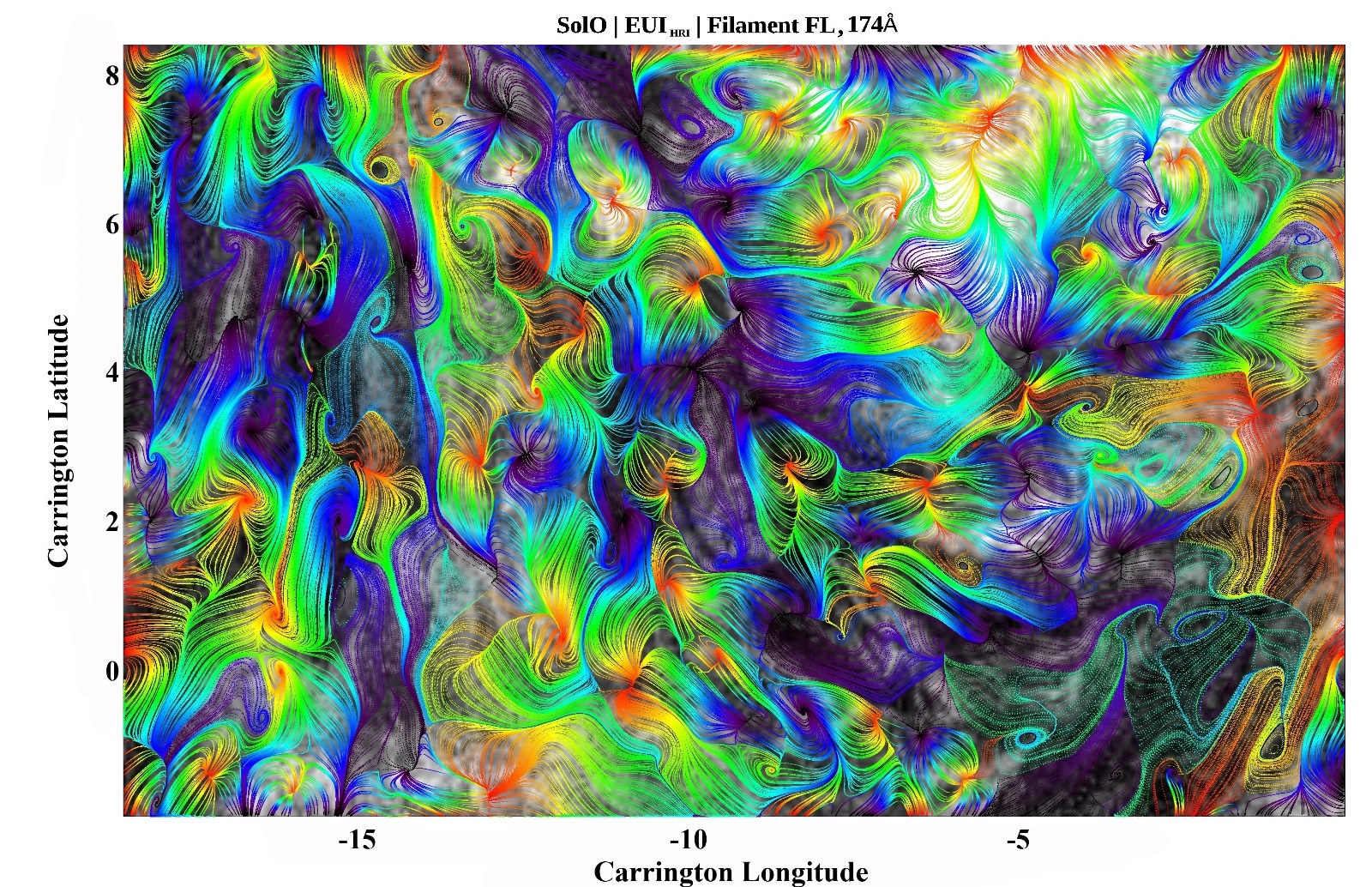}
	\includegraphics[width=0.45\linewidth]{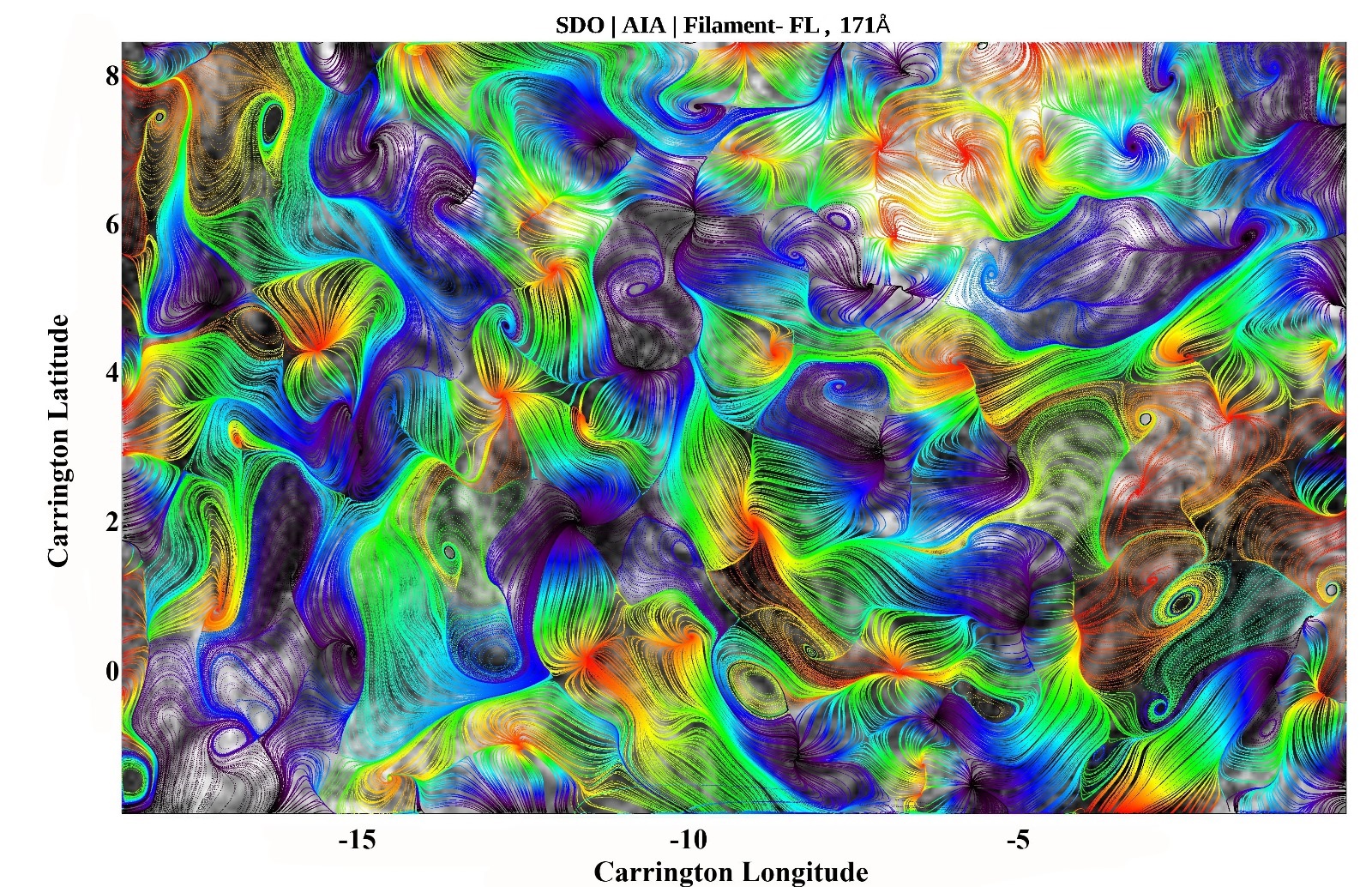}
	\includegraphics[width=0.45\linewidth]{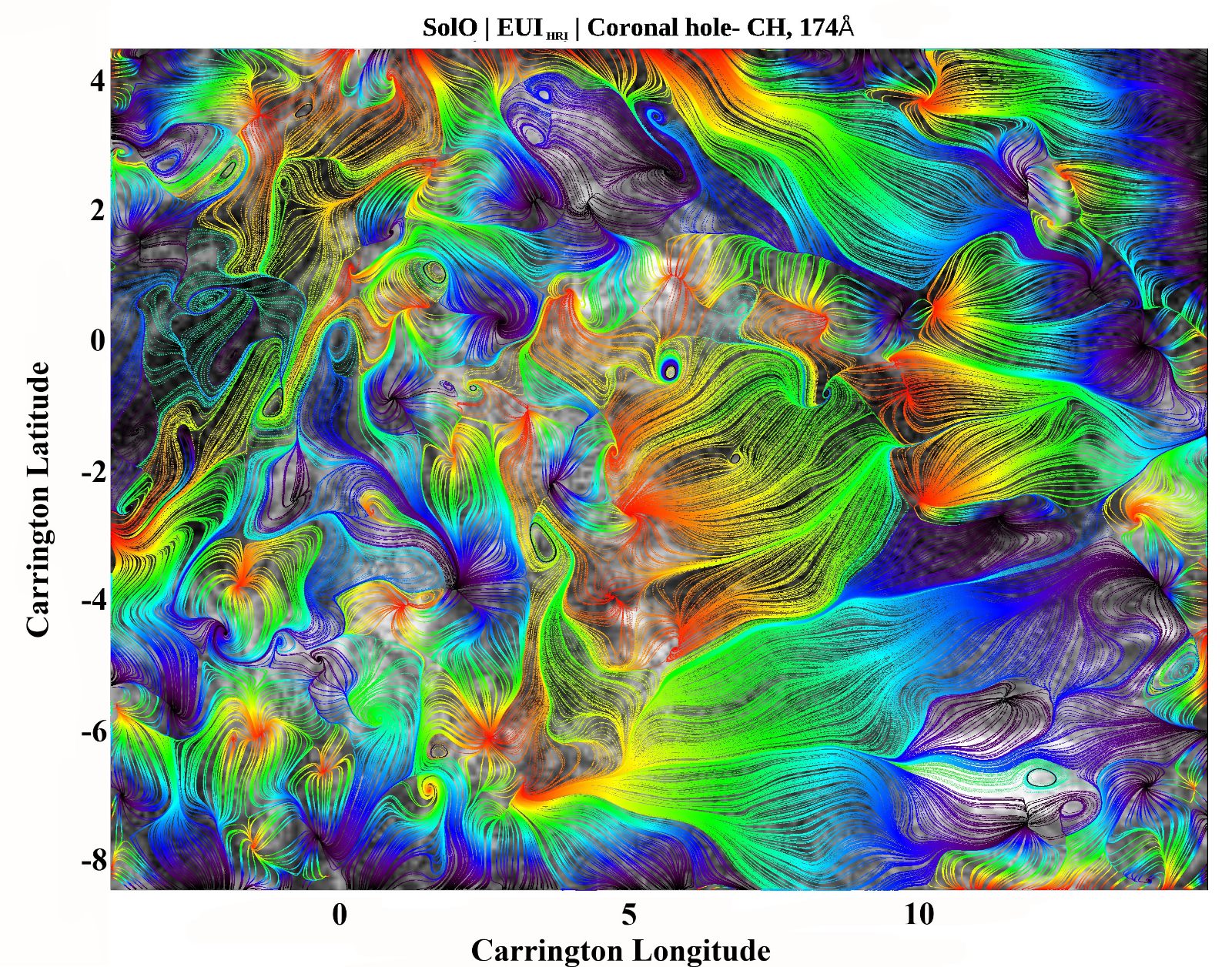}
	\includegraphics[width=0.45\linewidth]{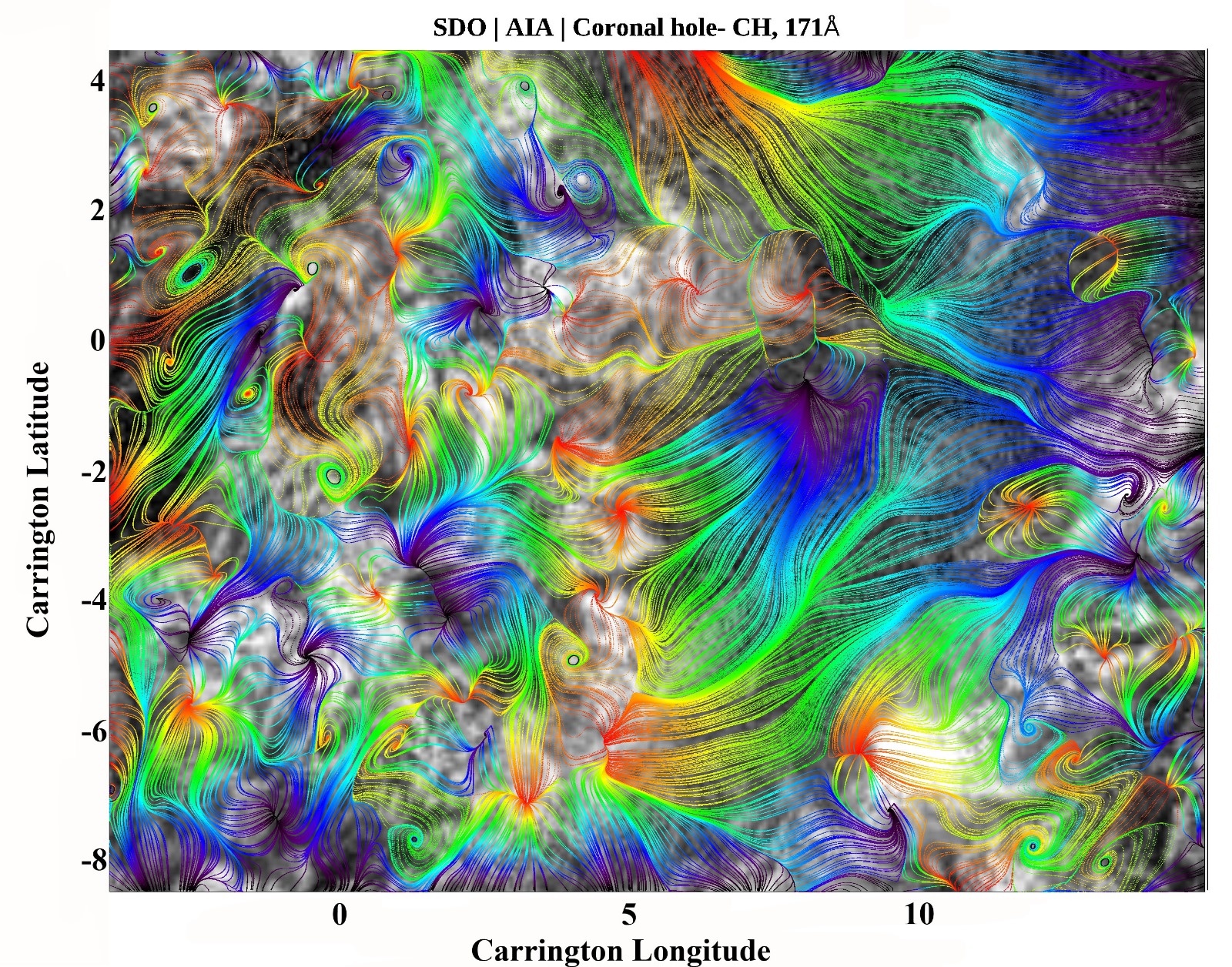}
	\caption{Greater detail of selected subregions showing QS (top), part of a filament (middle), and the CH (bottom). }
	\label{cuts}
\end{figure*}

The top row of Figure \ref{cuts} shows greater detail of a QS region nestled within the U-shaped filament. There are clearly-formed cells that appear in the vector maps of both instruments. For example, the source (red) cell centred approximately at (2,-11), and labelled in the top panels, clearly exists in both instruments, appearing as a point source in AIA but as two closely-separated point sources in HRIEUV. This is an example of a topological difference that we believe is most likely due to the different spatiotemporal resolutions, sensitivity, and temperature sensitivity of the instruments. We would expect the magnetic topology to vary between the different heights of emission viewed by both instruments. The cell in HRIEUV at coordinates (-2,-13) has an elongated sink (blue) region at it's centre. The flow lines converge from all directions into this elongated sink. In AIA, this feature seems shifted by half a degree in longitude to the east (to the left). This pattern of consistent features shifted by a small factor in coordinates between the instrument is likely due to two main reasons. (1) The global heights to the spherical surface for the mapping to Carrington coordinates may not be suitable for all regions. (2) The topology of the magnetic field may change between the dominant height layer observed by AIA and HRIEUV.

The U-shaped filament structure is dominated by blue coloured lines, indicating that PDs end their propagation within the filament. The middle row of Figure \ref{cuts} shows a closer view of part of a filament located at the top left corner of Figure \ref{6.2} . The fieldlines are generally aligned along the filament spine with a predominant single direction of motion. Thus PDs tend to propagate from closely-neighbouring QS regions into the filament, at which point they propagate along the filament in a consistent direction. This is seen in both instruments, although there are some differences in the actual detail at smaller scales, for the same reasons as listed above. 

The equatorial CH shows the interesting feature of very long coherent velocity fieldlines bridging across from one side of the CH to the other (from left to right), seen in more detail in the bottom row of Figure \ref{cuts}. These fieldlines originate from the QS to the east (left) of the CH, and extend across to the QS on the west (right) side. These are the longest coherent fieldlines throughout the ROI, and are structurally different to that of the QS and the filaments in that they do not exhibit a cell-like pattern. The intensity images of figure \ref{6.2} show a few faint, approximately linear, enhancements in intensity bridging across the CH with a similar configuration to the PD velocity fieldlines. The velocity field therefore must arise from PDs traversing along a QS magnetic field which bridges across the CH. This is contrary to the expected `open' magnetic field topology of CHs. The intensity images show only a few loops bridging across the CH, whilst the velocity field suggests that this is a general configuration across the whole CH.

\subsection{Speed distributions}
\begin{figure*}
	\centering
	\includegraphics[width=0.45\linewidth]{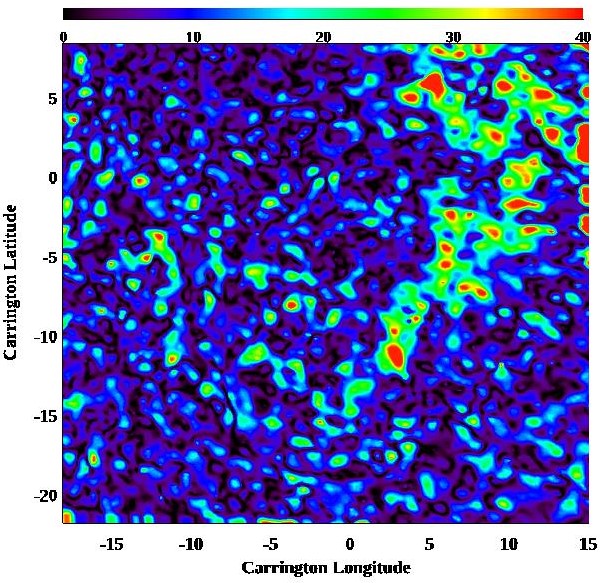}
	\includegraphics[width=0.45\linewidth]{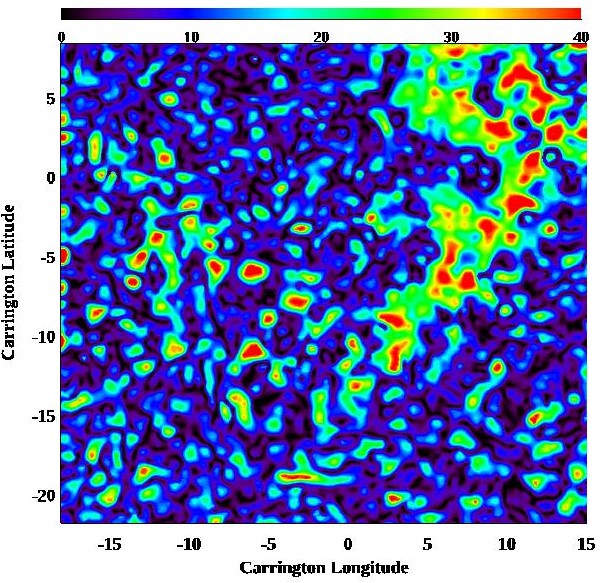}
	\caption{TNOF speeds over the ROI for HRIEUV (left) and AIA (right). Both maps share the same colour scale as shown in the colour bars above each plot.}
	\label{6.3}
\end{figure*}

Figure \ref{6.3} maps the TNOF speeds for both instruments. The CH appears as a broad region of higher speed in the 20-40\kms\ range for both instruments. Outside of the CH there are only isolated regions of higher speeds, with most regions showing values of below 10\kms\, or close to the most probable speed of around 5\kms. Speeds in the filament channel do not show the coherence of the CH, although there are clusters of higher speeds distributed along the boundaries of the southern `U-shaped' bend of the filament. Overall, our optical velocities are in the expected range of slow MA waves and they are in line with previous reports (e.g. \cite{nakariakov_2011_slow, zhao_2025_observed, demoortel_2014_observational, gupta_2012_spectroscopic}).   

\begin{figure*}
	\centering
	\includegraphics[width=0.32\linewidth]{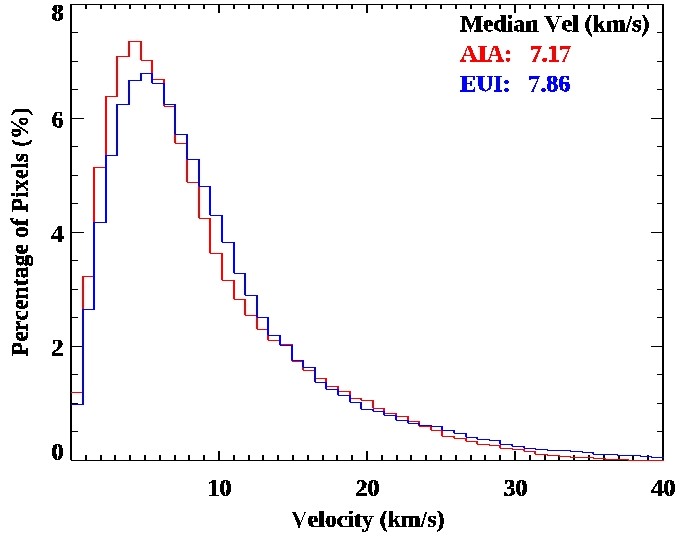}
	\includegraphics[width=0.32\linewidth]{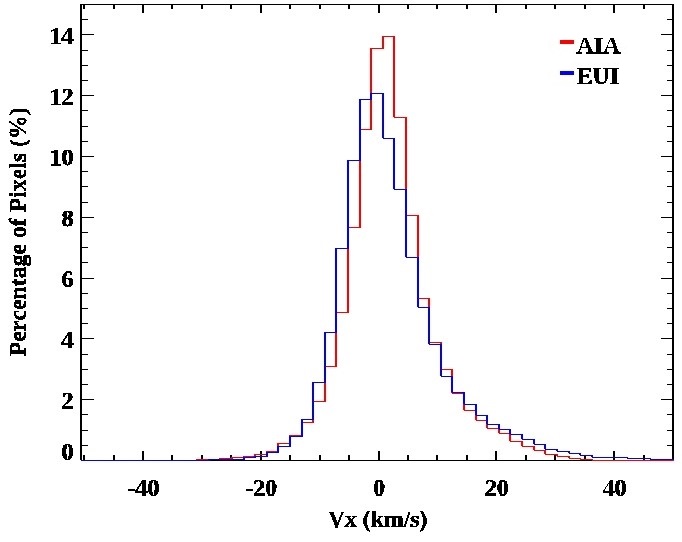}
	\includegraphics[width=0.32\linewidth]{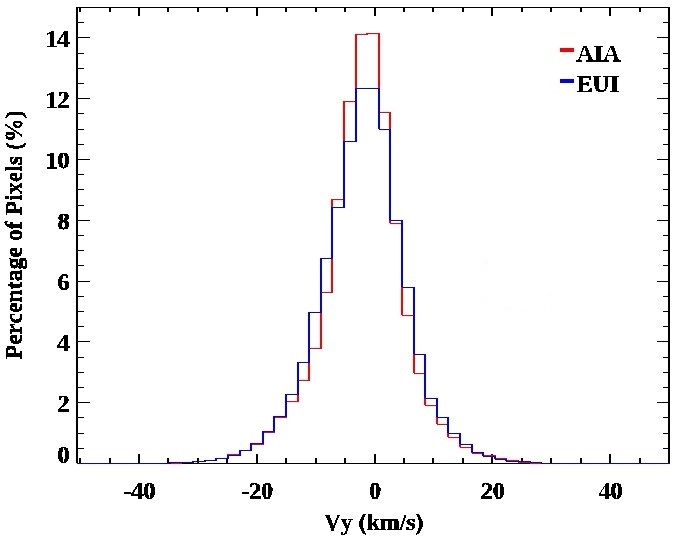}
	\caption{Left panel: The distribution of TNOF speeds across the whole ROI for AIA (red) and HRIEUV (blue). Middle and right panels: The $x$ (middle) and $y$ (right) components of velocity for both instruments.}
	\label{histocombined}
\end{figure*}

The left panel of figure \ref{histocombined} compares the distribution of speeds across the whole ROI for both instruments. The distributions are very similar. Both extend to velocities up to approximately 40\kms. AIA has a mean of 9.02\kms\ and HRIEUV is 10.14\kms. The most probable speed for AIA is 4.0\kms, compared to 4.8\kms\ for HRIEUV. The middle and right panels of Figure \ref{histocombined} compare the AIA and HRIEUV distribution of the $x$ (longitudinal) and $y$ (latitudinal) components of velocity across the whole ROI. The most probable $V_x$ speed is 1.5\kms\ for AIA and -1.6\kms\ for HRIEUV, and for $V_y$ both AIA and HRIEUV have most probable speeds of -1.4\kms. The small difference in $V_x$ could be due to the optimal height for the conversion to spherical coordinates, although this should also lead to a small difference in the $y$ component. A more subtle effect may arise from an early step in the method in which successive image frames are aligned by a global translational shift in $x$ and $y$. This global alignment best fits successive images, but it may not be accurate across the entire ROI because different areas should shift by different quantities (due to the spherical geometry and differential rotation). The same effect is present in both AIA and HRIEUV, but is exaggerated in AIA because of the faster apparent rotation of the Sun due to the spacecraft orbital dynamics. This effect will always be most apparent in the longitudinal component of the velocity. 

\begin{figure*}
	\centering
	\includegraphics[width=0.33\linewidth]{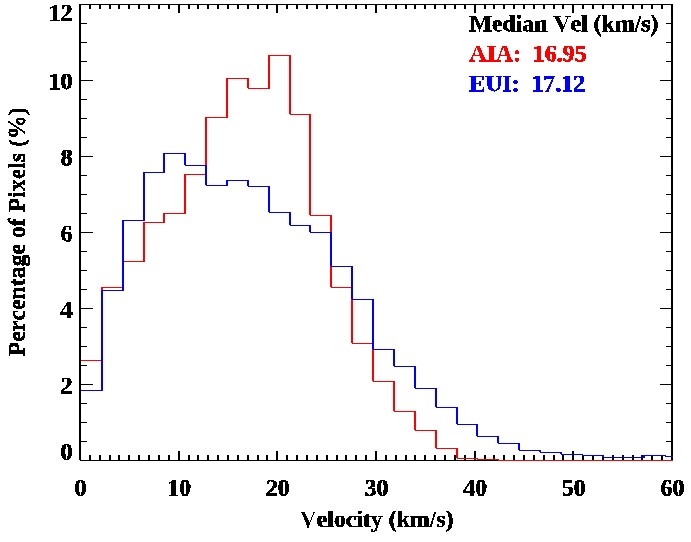}
	\includegraphics[width=0.33\linewidth]{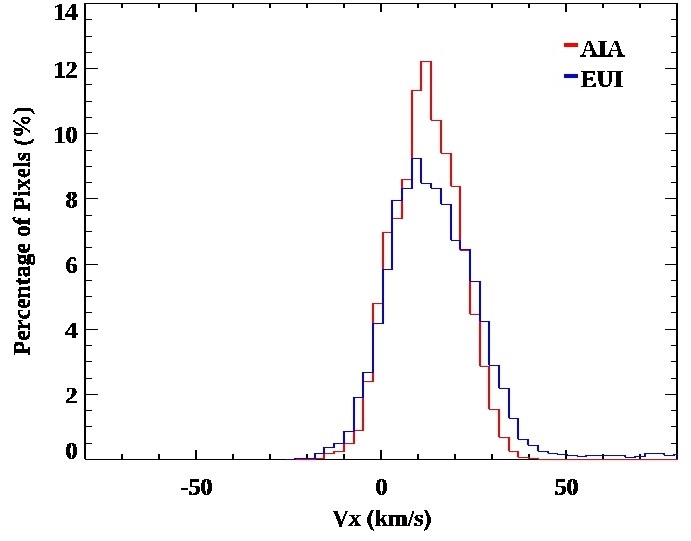}
	\includegraphics[width=0.33\linewidth]{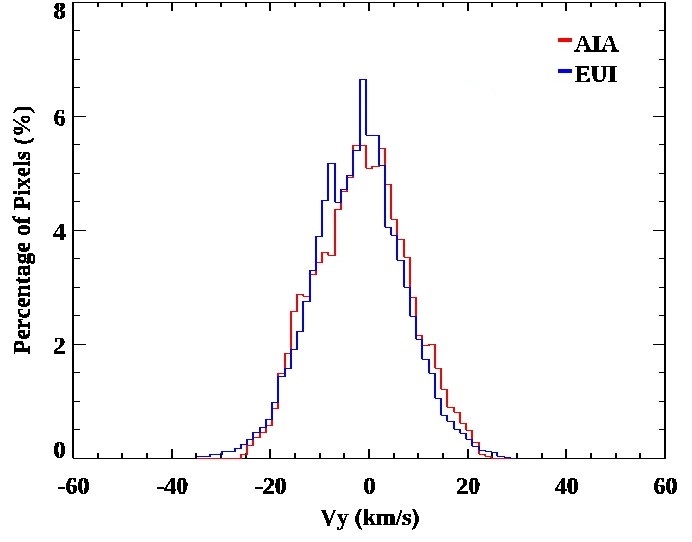}
	\caption{Left panel: Histogram of speeds (left), and $x$ (middle) and $y$ (right) components of velocity CH region for AIA (red) and HRIEUV (blue).}
	\label{fig:ch_vel}
\end{figure*}

\begin{figure*}
	\centering
	\includegraphics[width=0.32\linewidth]{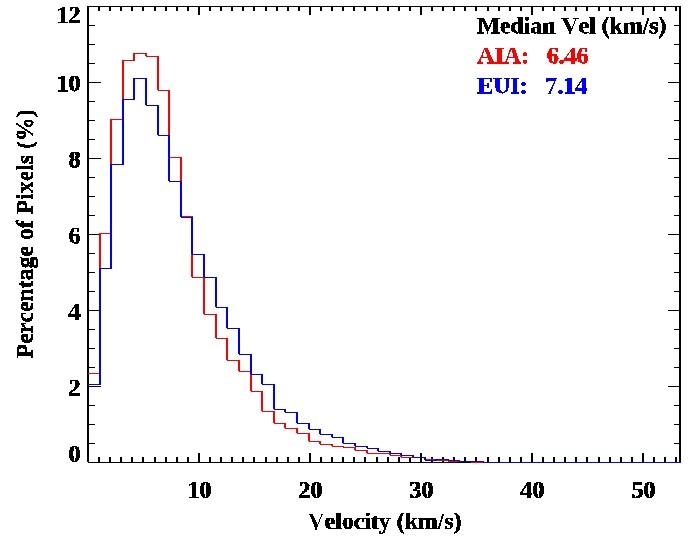}
	\includegraphics[width=0.32\linewidth]{figures/new/velocity_new_vx_1.jpg}
	\includegraphics[width=0.32\linewidth]{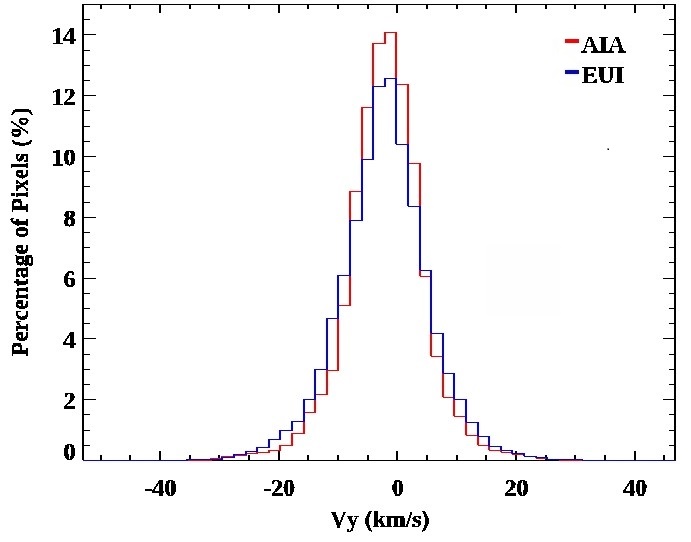}
	\caption{Left panel: Histogram of speeds for the filament region for AIA (red) and HRIEUV (blue). Middle and right panels: The $x$ (middle) and $y$ (right) components for both instruments}
\label{FL}
\end{figure*}

Figure \ref{FL} shows histograms of the filament region speeds for AIA (red) and HRIEUV (blue). Both distributions agree very closely. Speeds peak at values near 5\kms\ and decline to a maximum around 25\kms. These speeds are considerably lower than the CH. The $x$ and $y$ speed components shown in the middle and right panel agree well, with the largest difference seen in the $x$ component of the middle panel. Here, HRIEUV is shifted to slightly higher positive values. Similar to the CH, this could be due to the different sensitivity of the instruments and other effects, but could also be due to the projected true velocities seen from the different longitudinal viewpoints.

\begin{figure*}
\centering
\includegraphics[width=0.32\linewidth]{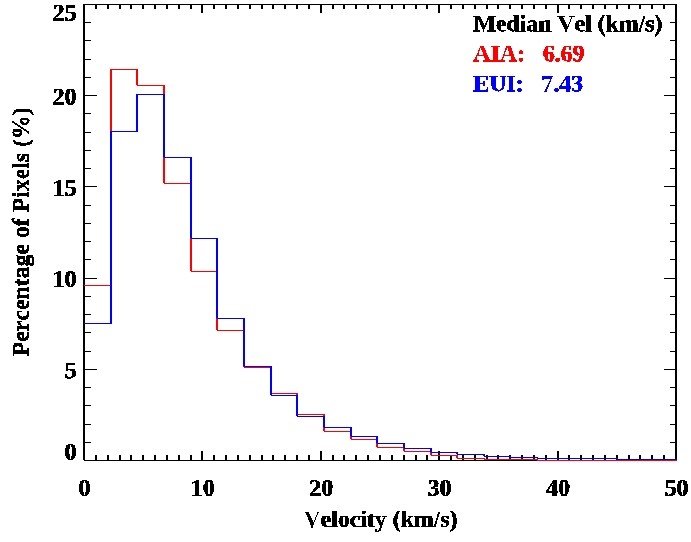}
\includegraphics[width=0.32\linewidth]{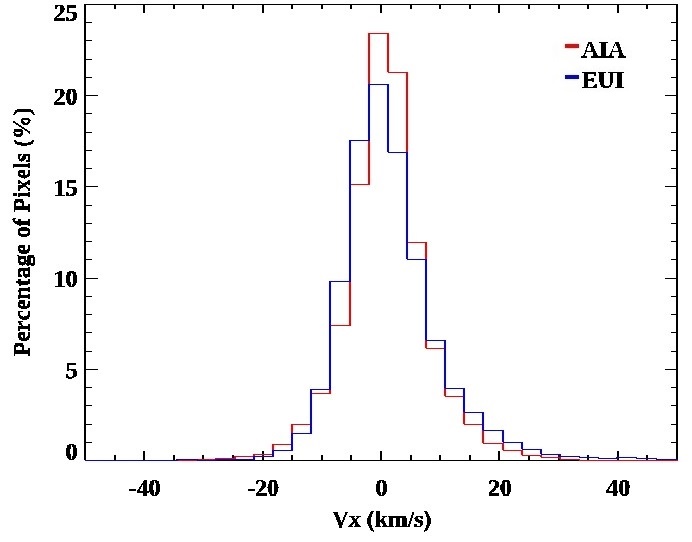}
\includegraphics[width=0.32\linewidth]{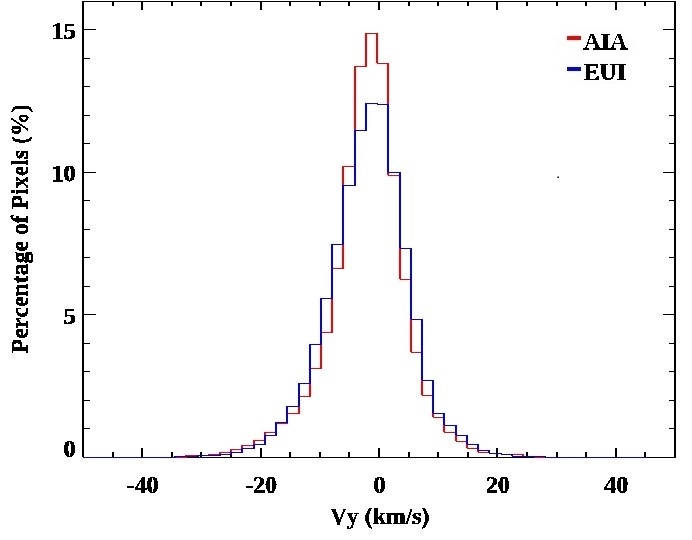}
\caption{Left panel: Histogram of speeds for the quiet sun region for AIA (red) and HRIEUV (blue). Middle and right panels: The $x$ (middle) and $y$ (right) components for both instruments}
\label{QS}
\end{figure*}

Figure \ref{QS} shows histograms of speed in the quiet sun regions (i.e. not belonging to the CH or filament) for AIA (red) and HRIEUV (blue). Both distributions agree very closely. Speeds peak at values AIA is 2.30\kms and 4.60\kms for HRIEUV and decline to a maximum around 30\kms. The mean speed for AIA is 8.23\kms and 9.25\kms. These speeds are considerably lower than the CH and a little lower than the filament. The $x$ and $y$ speed components shown in the middle and right panel agree well, with the small difference seen in the $x$ component of the middle panel. Here, HRIEUV is shifted to slightly higher positive values. Similar to the previous histograms, this could be due to the different sensitivity of the instruments and other effects, but could also be due to the projected true velocities seen from the different longitudinal viewpoints.

\section{Discussion}
 
The CH is a large region of consistently higher speeds than the rest of the ROI. We believe this is linked to the system of linear magnetic loops bridging across the CH which may be conducive to such high speeds compared to the smaller systems of loops that exist in the QS. Alternatively or in addition, these loops are possibly rooted in QS network regions of higher magnetic field strength which may drive faster PDs, although other loop systems in the QS are rooted in photospheric network regions of similar strength and do not show such high speeds. 

\begin{figure}
	\centering
	\includegraphics[width=1.0\linewidth]{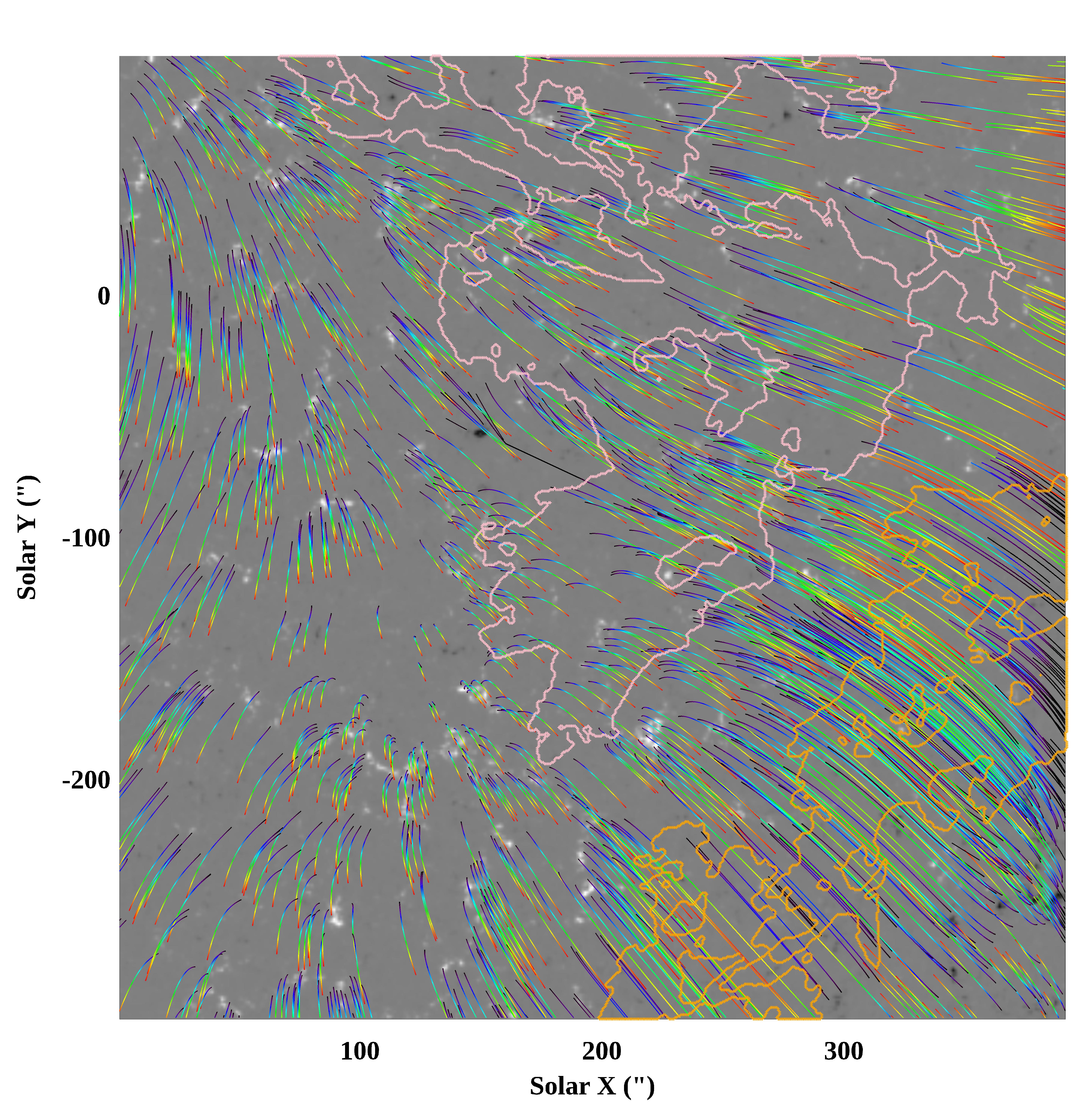}
	\caption{Magnetic fieldlines from a PF model of the HMI observed photospheric field for a region encompassing the CH. The background image is the HMI observation, with positive (negative) regions in white (black). The CH is bounded in pink and part of the filament channel in orange. The colour along a fieldline represents the height above the photosphere, ranging from black (low height) through blue, green, yellow and red for increasing height to the limit of the model volume at $\sim 0.1$\Rs. The model field is calculated using a Green's function approach, with the region of the photosphere used for the extrapolation extending beyond that shown in this figure.}
	\label{pfss}
\end{figure}

Figure \ref{pfss} shows fieldlines from a potential field (PF) model overlaid on a HMI magnetogram with the model limited to a region encompassing the CH. The model is created using Green's function and considers all pixels from a magnetogram region with margins extending beyond the model volume. HMI shows that photospheric positive polarities are slightly dominant across this whole region (CH and QS), with a mean field of 3.2G. The photospheric field strength within the CH is slightly lower than the surrounding QS region (8.3G compared to 9.0G), and has a higher excess positive flux (mean field of 5.0G compared to 2.9G for the QS). Whilst some local polarities of opposite signs are linked by smaller loops, there are systems of overlying fieldlines that are long and link more distant regions including fieldlines that extend across the CH. This is in general agreement with the TNOF velocity map of figure \ref{6.2}. Although a PF extrapolation of this limited region containing generally weak field (i.e. with higher relative uncertainty in the photospheric measurement) is not likely to give a wholly accurate model of the coronal field, Figure \ref{pfss} gives support to our interpretation of the coronal field above the CH. 

This CH is old (has persisted for several solar rotations prior to these observations), and is narrow in longitudinal extent, as is typical of small equatorial CHs. We believe that the closed magnetic field from the surrounding QS regions may, through gradual small-scale reconnection with the open field, evolve so it eventually bridges across the CH. This implies that the CH may not possess open field. The question then arises why it appears so clearly as a dark, low-temperature region in the EUV. The properties of the plasma are typical of a CH, yet the magnetic field properties are not.

The left panel of Figure \ref{fig:ch_vel} shows significant differences between the HRIEUV and AIA speed distributions within the CH. This is possibly due to the spatiotemporal resolution and sensitivity as described previously. Another possibility is the effect of the projected true PD velocities in the corona to the different instrument's image planes. Such a projection effect would be difficult to see in the QS or filament due to the smaller-scale structure and complexity but should be most apparent in a region such as this CH, where there is a system of coherent linear lines of motions across a broad region. The CH is closer to the disk center for AIA compared to HRIEUV, and subject to less projection effects. Furthermore, the projection effect is greatest along the longitudinal dimension compared to the latitudinal. This argument supports the greater difference in the instrument's mean speeds in the longitudinal ($x$) direction compared to the latitudinal ($y$) direction. 

Another effect could be the different height of peak emission of the two instruments. Previous works using TNOF have shown clear topological differences in the velocity fields from different AIA channels which are dominated by different temperatures/heights in the atmosphere. For example, the 304\AA\ channel shows a smaller-scale cell-like pattern compared to the 171\AA\ and 193\AA\ channels. The 193\AA\ channel in particular shows the largest coherent cell-like structures. We would expect the magnetic topology, as well as the PD speeds, to change between the 4 Mm AIA 171\AA\ height and the 11.4Mm HRIEUV 174\AA\ height. The challenge is to distinguish these real topological features from other effects, and this will require further development.

As seen in Figure \ref{6.2} the dominant colour of the velocity fieldlines within the FC is blue to black, which means that most PD velocity vectors start near the filament channel boundary and end within the filament. This suggests that PD activity in filaments are dominated by external drivers, although the core of the filament is likely cool and optically thick, and will not be seen in our observations. These are elongated field lines that tend to align approximately along the filament spine. This pattern of PD propagation is consistent with the expected large-scale magnetic field topology of filaments, where fieldlines rooted close to the filament channel enter the flux tube of the filament and become aligned with the overall flux tube axis. The lower right portion of Figure \ref{pfss} shows a large arcade of PF fieldlines bridging over part of the filament channel (outlined in orange), with the fieldlines approximately perpendicular to the alignment of the channel. This is in considerable disagreement with the filament-aligned PD velocity fieldlines and shows the inability of a PF model to accurately depict the highly non-potential magnetic field of a filament. 

Regions outside of the CH and FC which we have labelled QS are thought to consist of systems of loops lying low in the corona. We have shown previously that the cell-like patterns covering QS regions are closely aligned with the photospheric network, with the network generally associated with PD sources. This is in agreement with the generally-accepted model of the QS loop footpoints being embedded within the network. Whilst the TNOF velocity vector fields contain topological differences at smaller scales, the distribution of QS speeds are very similar. This suggests either that PD speeds are similar over a range of low coronal heights in the QS, or that the motions are most apparent in the temperature band that is common to both instruments. To resolve such questions will require future instruments with a finer temperature resolution.

\section{Conclusions and future work}

Our comparison of the PD velocity fields and speeds gained from HRIEUV 174\AA\ and AIA 171\AA\ observations show an overall excellent agreement which increases confidence that the TNOF method is accurately estimating the motions of PDs in the low corona. We can interpret differences in PD speeds and velocity vector maps in terms of the real variation expected with height in the corona, with the different temperature sensitivity of the instruments viewing different layers. However, such an interpretation is limited because it is currently impossible to distinguish this from other effects such as differences in instrumental spatiotemporal resolution, projection effects due to the different viewpoints, and uncertainties in the results. 

To enable the comparisons in this work, we remap to Carrington coordinates and optimise the heights of the spherical surfaces used as a coordinate basis for each instrument, where we find optimised heights of 11.4Mm for HRIEUV 174\AA\ and 4 Mm for AIA 171\AA. This is a large difference in height for a relatively small difference in temperature sensitivity, which implies that the temperature height gradient is gradual in the quiet low corona. This method can be readily applied to other suitable EUI and AIA datasets. We wish to apply this approach to smaller regions, thus mapping the height of dominant emission over different types of coronal structures. Currently our attempts at this localised mapping give unstable results thus further development is required.

The TNOF velocity vector map shows that the CH has a system of low-lying closed magnetic field bridging from east to west, rooted in nearby QS regions. This is supported by a PF magnetic model extrapolation of the observed photospheric field. This is in contradiction to the general CH model of open magnetic field. The CH is a region of consistently higher speeds than the QS and FC which suggests that longer quasi-linear magnetic loops allow higher PD speeds compared to the smaller loop systems of the QS and FC.

The FC velocity vector field is a configuration in agreement with the generally accepted model of a filament's highly non-potential tubular magnetic field, which the PF magnetic model fails to replicate. We show also evidence that PD activity in filaments arises from external drivers. In the QS, we see a similar cell-like structure to the velocity field which we have shown previously to be aligned with the photospheric network. PD speed distributions across both QS and FC regions are very similar between HRIEUV and AIA, suggesting either that PD speeds do not vary much with height or that the instrument's observations are dominated by the same layer.

A long-term goal is to use different spacecraft viewpoints to enable a three-dimensional constraint on PD velocity fields, although this is challenging. When the longitudinal separation is suitable, applying TNOF to the same region from both perspectives may allow us to assess how velocity fields vary with viewing angle. This will help determine the extent to which differences reflect magnetic field alignment, particularly above network boundaries, and whether this approach can provide practical constraints on field connectivity despite the uncertainties in height and projection. Another approach can include comparing spectral Doppler measurements and numerical simulations. Depending on their nature, the PDs may not always show a Doppler shift. Yet, the source and sink regions may still carry significant flows which are field-aligned, complementing the TNOF PD velocity fields which show the plane-of-sky component. This type of analysis would offer a novel and valuable insight into the magnetic field topology.

\begin{acknowledgements}
   We acknowledge support from Aberystwyth University through the Aberdoc PhD scholarship scheme and the President's award. We acknowledge the support of the Royal Observatory of Belgium’s Solar Physics team through the 2024 Guest Investigator Program, which supported this research via access to the EUI data PI team in Brussels. We thank Sarah Willems and Koen Stegen for their help with IDL and Solar Soft. H. Morgan conducted work for this study under STFC grant WT414417-01. N.Narang acknowledges funding from the Belgian Federal Science Policy Office (BELSPO) contract B2/223/P1/CLOSE-UP.  The SDO data used in this paper is courtesy of NASA/SDO and the AIA, EVE, and HMI science teams. Solar Orbiter is a space mission of international collaboration between ESA and NASA, operated by ESA. The EUI instrument was built by CSL, IAS, MPS, MSSL/UCL, PMOD/WRC, ROB, LCF/IO with funding from the Belgian Federal Science Policy Office (BELSPO/PRODEX PEA 4000112292 and 4000134088); the Centre National d’Etudes Spatiales (CNES); the UK Space Agency (UKSA); the Bundesministerium für Wirtschaft und Energie (BMWi) through the Deutsches Zentrum für Luft- und Raumfahrt (DLR); and the Swiss Space Office (SSO). This research used the Heliophysics Event Knowledge database and the ESA JHelioviewer. 
\end{acknowledgements}

\bibliography{Bib}

\bibliographystyle{aasjournal}

\end{document}